\documentclass[lettersize,journal]{IEEEtran}
\usepackage{amsmath,amsfonts,amssymb}
\usepackage[linesnumbered,lined,boxed,commentsnumbered, ruled]{algorithm2e}
\usepackage{array}
\usepackage{textcomp}
\usepackage{stfloats}
\usepackage{url}
\usepackage{verbatim}
\usepackage{graphicx}
\usepackage{cite}
\usepackage{bm}
\usepackage{subfigure}
\usepackage{multirow}
\usepackage{color}
\begin{document}

\title{Wasserstein Adversarial Learning based Temporal Knowledge Graph Embedding}

\author{Yuanfei Dai, Wenzhong Guo, ~\IEEEmembership{Member,~IEEE,} and Carsten Eickhoff
		
\thanks{Yuanfei Dai is with the Department of Computer Science and Technology, Nanjing Tech University, Nanjing 211816, China. (email: daiyuanfei@njtech.edu.cn).}
\thanks{Wenzhong Guo is with the Department of Computer and Big Data, Fuzhou University, Fuzhou 350108, China. (email: guowenzhong@fzu.edu.cn).}
\thanks{Carsten Eickhoff is with the Center for Biomedical Informatics, Brown University, Providence 02912, USA. (email: carsten@brown.edu).}}
\maketitle

\begin{abstract}
Research on knowledge graph embedding (KGE) has emerged as an active field in which most existing KGE approaches mainly focus on static structural data and ignore the influence of temporal variation involved in time-aware triples. In order to deal with this issue, several temporal knowledge graph embedding (TKGE) approaches have been proposed to integrate temporal and structural information in recent years. However, these methods only employ a uniformly random sampling to construct negative facts. As a consequence, the corrupted samples are often too simplistic for training an effective model. In this paper, we propose a new temporal knowledge graph embedding framework by introducing adversarial learning to further refine the performance of traditional TKGE models. In our framework, a generator is utilized to construct high-quality plausible quadruples and a discriminator learns to obtain the embeddings of entities and relations based on both positive and negative samples. Meanwhile, we also apply a Gumbel-Softmax relaxation and the Wasserstein distance to prevent vanishing gradient problems on discrete data; an inherent flaw in traditional generative adversarial networks. Through comprehensive experimentation on temporal datasets, the results indicate that our proposed framework can attain significant improvements based on benchmark models and also demonstrate the effectiveness and applicability of our framework. 
\end{abstract}


\section{Introduction}
\label{section1}
\IEEEPARstart{K}{nowledge} graphs (KGs), also called knowledge bases (KBs), are employed for gathering and organizing distributed human knowledge and information in a graph structure where nodes represent entities and edges indicate relations. In recent years, a variety of large-scale KGs such as DBpedia~\cite{dbpedia}, NELL~\cite{NELL} and Wikidata~\cite{wikidata} have been very successfully applied for many natural language processing (NLP) tasks including machine reading~\cite{yang2017leveraging, qiu2019machine}, question answering~\cite{huang2019knowledge, 9199272}, and information retrieval~\cite{dietz2018utilizing,9559716}. A typical knowledge graph is composed of various triple facts ${(h, r, t)}$, in which $h$ and $t$ denote head and tail entities respectively, and $r$ indicates a relationship from $h$ to $t$, e.g., (\emph{Albert Einstein, has won price, Nobel Prize in Physics}).

Unfortunately, issues of data sparsity and computational complexity have intensified with the increasing amount of information contained in large-scale KGs, which could cause management and manipulation difficulties. In order to deal with these problems, knowledge graph embedding (KGE) has been proposed and gained substantial attention~\cite{transE2013, transH2014, transR2015, 9416312}. The main idea of a KGE model is to represent both entities and relations in low-dimensional feature spaces so as to condense data space and simplify calculation while preserving the inherent nature of the original graph.

\begin{figure}[t]
	\centering
	\includegraphics[width=8cm]{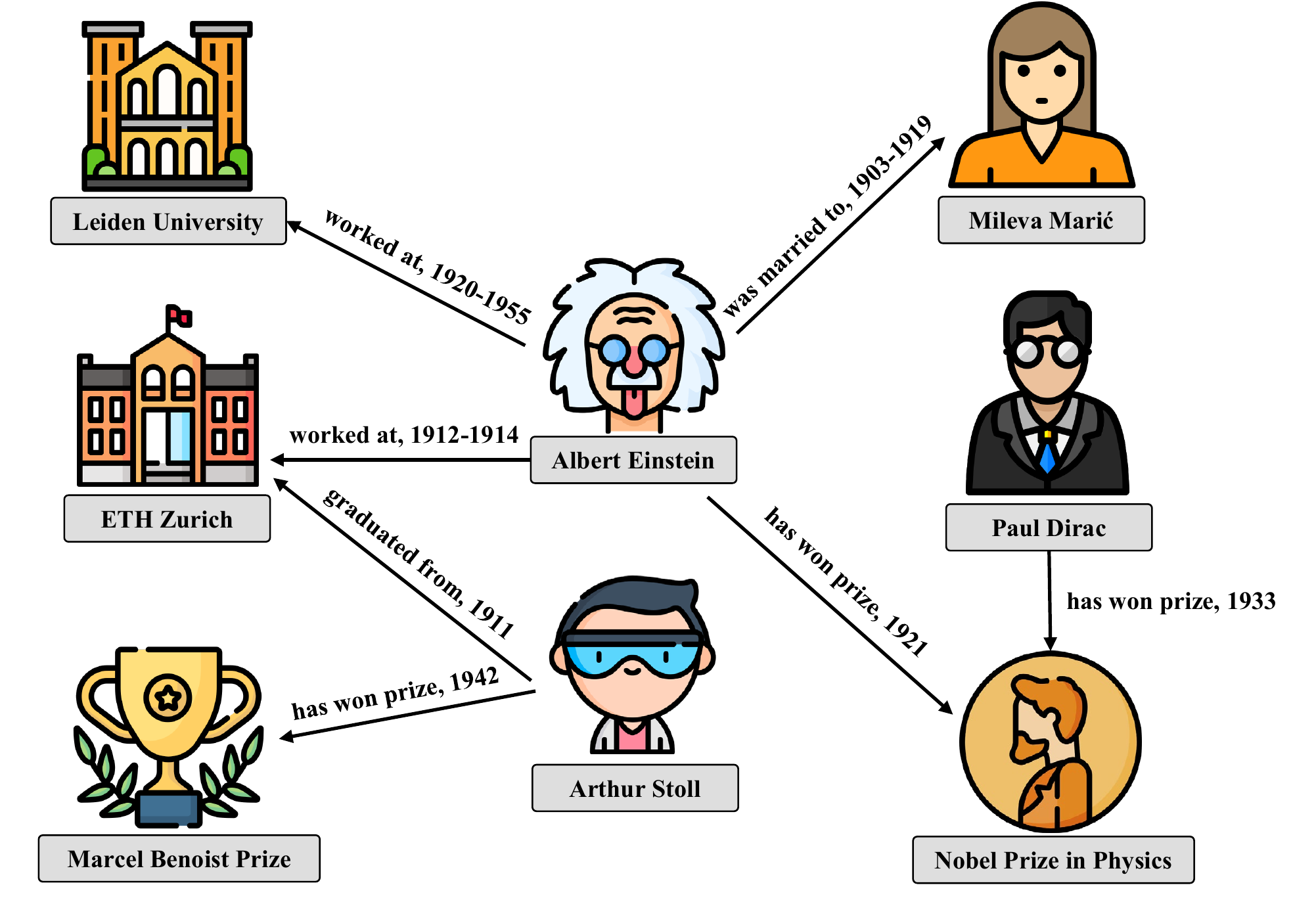}
	\caption{A simple temporal knowledge sub-graph to illustrate the necessity of considering time-aware information on KGs. Any fact can only be established at a specific timestamp.}
	\label{Figure1}
\end{figure}

Existing KGE approaches mainly focus on static knowledge graphs based on the assumption that the contained facts are statically true and will not change over time. However, in real-life situations, most objective facts are only valid during a certain time point or period, in other words, KGs in the real world tend to be dynamic and the authenticity of triples evolves over time. For instance, as displayed in Figure \ref{Figure1}, a sentence ``\emph{Albert Einstein has won the Nobel Prize in Physics in 1921.}'' can be constructed as one triple (\emph{Albert Einstein, has won price, Nobel Prize in Physics}), and this fact was only correct in 1921. Since existing static KGE methods solely learn from time-agnostic triple facts and ignore the potentially beneficial temporal information, it is important to provide an embedding approach for dynamic KGs.

To achieve this goal, various temporal knowledge graph embedding (TKGE) models have been provided to incorporate temporal information in their embedding vectors. They usually calculate a hidden embedding vector for each timestamp and extend the final score function by absorbing this representation as well as entity and relation embeddings. TKGE methods proved that they outperform traditional KGE models on temporal KGs. However, such models generate negative facts through a uniform negative sampling strategy~\cite{transE2013} that replaces head or tail entities in a positive triple with a different one from the entity candidate set in which all entities share the same sampling weights. This simple sampling scheme usually introduces only very limited benefits to the performance of the learned embedding model and can even delay model convergence~\cite{schroff2015facenet, hermans2017defense}.

In recent years, generative adversarial networks (GANs)~\cite{goodfellow2014generative} have become a popular research direction owing to their powerful generalization and representation capabilities. They have not, however, yielded satisfying results in many natural language processing (NLP) tasks as the original GANs perform well on continuous data, but cannot be directly deployed on discrete data due to issues of vanishing gradients. To address this drawback, policy gradient, a type of reinforcement learning (RL) algorithm, was introduced to replace the existed back propagation, this new strategy was then applied in traditional KGE models to generate more plausible negative samples to refine a model's performance~\cite{wang2018incorporating, kbgan2018}. Although these architectures have been proven effective, high-variance optimization processes require a large number of computational resources while brittle hyper-parameters increase instability of the already difficult-to-train GANs during training time.

In this paper, we propose a robust temporal knowledge graph embedding framework based on adversarial learning. The proposed framework utilizes a generator to construct more plausible negative triples. Afterwards, these negative samples are incorporated with positive triples and jointly fed into a discriminator to yield the embedding vectors which form the final representation of the TKGE model. Moreover, we introduce Gumbel-Softmax relaxation and the Wasserstein distance to handle the issue of vanishing gradients on discrete data without policy gradient mechanisms. The main contributions in this paper are:

\begin{itemize}
	\item Learning the embedding vectors for temporal knowledge graphs is a scarcely explored research field as most existing KG representation approaches solely train models from triples without time-unknown facts. In this paper, we propose a new TKGE embedding framework to allow embedding models acquiring representations by incorporating temporal information.
	
	\item To the best of our knowledge, we are the first to introduce generative adversarial learning to TKGE. The generator is capable of constructing more plausible negative samples and the discriminator utilizes these positive and negative triple facts to train the final TKGE model. In addition, our proposed framework is remarkably extensible and can be applied to most existing embedding models.
	
	\item We comprehensively evaluate the robustness and effectiveness of our proposed framework on a link prediction task on five TKG benchmark datasets. The experimental results show that our framework can effectively improve the performance of the original TKGE models.
\end{itemize}

The remainder of this paper is organized as follows. In Section \ref{section2}, we
introduce several existing static and dynamic knowledge graph embedding models and briefly highlight the connections and differences between them. Section \ref{section3} illustrates the proposed overall adversarial framework and its training procedure in detail. Section \ref{section4} delineates experimental details, including benchmark datasets, experimental parameter initialization settings and results. We provide a qualitative comparison and discussion between the results obtained via our framework and the original methods in Section \ref{section5}. Finally, concluding remarks are deliberated in Section \ref{section6}.

\section{Related Works}
\label{section2}
In this section, we begin by introducing several basic notations and corresponding explanations that will be used in the remainder of this paper. Afterward, we supply a general definition of the knowledge graph representation learning problem and introduce a representative range of embedding models.

Ordinary lowercase letters represent scalars, bold lowercase letters represent vectors and bold uppercase letters represent matrices. Given a knowledge graph consisted of a set of triple $\mathcal{G} = <s, p, o>$, where $s$ and $o$ are both in entity set $\mathbb{E}$ and $p$ belongs to relation set $\mathbb{R}$. KG embedding aims to transform each entity and relation into a low-dimensional feature space. A large number of KGE methods have been proposed. These works can be roughly categorized into two branches: conventional static knowledge graph embedding methods and the emerging field of dynamic knowledge graph embeddings.

\subsection{Static knowledge graph embedding}
Mikolov et al.~\cite{mikolov2013distributed} proposed a word embedding algorithm by following the translation invariance principle that words with similar connotation should have similar representations. Inspired by this translation based idea, Bordes et al.~\cite{transE2013} extended the same principle to the knowledge graph, and proposed the TransE model. TransE interprets relations as translation vector connected vectors of head and tail entities, i.e., $\bm{e}_{s} + \bm{e}_{p} \approx \bm{e}_{o}$. The authenticity of each triple ($s, p, o$) is measured by a score function. The score implies the distance between $\bm{e}_{s} + \bm{e}_{p}$ and $\bm{e}_{o}$, and the function is shown as follows:
\begin{equation}
	f(s, p, o) = \left \| \bm{e}_{s} + \bm{e}_{p} - \bm{e}_{o} \right \| _{\ell_{1}/\ell_{2}}.
	\label{equ2}
\end{equation}

Here, $\left \| \cdot \right \| $ denotes the norm operation, $\ell_{1}$, $\ell_{2}$ are $L_{1}$-norm and $L_{2}$-norm,  respectively. Even though TransE achieves solid KGE performance, it struggles to address complex relations, such as $1-N$, $N-1$, and $N-N$. Therefore, other embedding models, such as TransH~\cite{transH2014}, TransR~\cite{transR2015}, TransG~\cite{transG}, etc., have been proposed.

Different from the above approaches, tensor factorization based KGE methods such as RESCAL~\cite{nickel2011three}, DistMult~\cite{yang2015embedding}, ComplEx~\cite{trouillon2016complex}, SimplE~\cite{kazemi2018simple} and so forth are another effective category for knowledge graph embeddings, in which each relation $p$ transforms into a latent semantic meaning matrix $\bm{M}_{p}$ and the score function $f(s, p, o)$ is formulated as:
\begin{equation}
	f(s, p, o) = \bm{e}_{s}^{\top}\bm{M}_{p}\bm{e}_{o},
\end{equation}
where $\bm{e}_{s}$, $\bm{e}_{o} \in \mathbb{R}^{d}$ denote entity embedding vectors. In recent years, deep neural network-based approaches have received considerable development. Several methods~\cite{conv2018,dettmers2018convolutional,balavzevic2019hypernetwork,xiao2019knowledge,li2021learning,li2021global} utilized feed-forward or convolutional neural networks (CNN) for scoring the authenticity of given triples.

\subsection{Dynamic knowledge graph embedding}
The above methods have yielded solid results in knowledge graph embedding applications. However, these traditional KGE models have limitations when processing temporal facts. As we illustrate in Figure \ref{Figure1}, the authenticity of objective facts is given in a distinct period and can change over time. To overcome this deficiency, some novel attempts are proposed to integrate temporal information for modeling dynamic knowledge graphs.

Leblay et al.~\cite{ttranse} extended the traditional TransE into a temporal embedding model TTransE via an additional embedding transformation: from timestamps to hidden representations. For each temporal fact $(s, p, o, t)$, time $t$ is also embedded in the same feature space as entities and relations. The final score
function is modified as:
\begin{equation}
	f(s, p, o, t) = \left \| \bm{e}_{s} + \bm{e}_{p} + \bm{e}_{t} - \bm{e}_{o} \right \| _{\ell_{1}/\ell_{2}}.
\end{equation}

HyTE~\cite{hyte} aimed to make entities have different distributed representations at different time points. It transforms time as a hyperplane, each embedding vector $\bm{e}_{s}$, $\bm{e}_{p}$ or $\bm{e}_{o}$ is projected on time-specific hyperplanes $\bm{w}_{t}$ as:
\begin{equation}
	\begin{array}{l}
		\overline{\bm{e}_{s}}=\bm{e}_{s}-\bm{w}_{t}^{\top} \bm{e}_{s} \bm{w}_{t} \\
		\overline{\bm{e}_{p}}=\bm{e}_{p}-\bm{w}_{t}^{\top} \bm{e}_{p} \bm{w}_{t} \\
		\overline{\bm{e}_{o}}=\bm{e}_{o}-\bm{w}_{t}^{\top} \bm{e}_{o} \bm{w}_{t}.
	\end{array}
\end{equation}

Intuitively, HyTE represents entities and relations to their corresponding time-specific spaces and utilizes the primitive TransE score function to distinguish the authenticity of given facts.

Garcia-Duran et al.~\cite{TAdistmult} introduced Long Short-Term Memory (LSTM) networks to obtain time-aware representations of relations. In their models, each timestamp can be decomposed into a sequence of temporal characters, a time-aware relation embedding vector $\bm{e}_{p,t}$ is calculated by feeding each element in this time sequence to an LSTM and taking its final output. Based on the previous model, the authors present two different improved versions TA-TransE and TA-DistMult, their score function are defined as follows:
\begin{equation}
	TA-TransE: 	f(s, p, o, t) = \left \| \bm{e}_{s} + \bm{e}_{p,t}- \bm{e}_{o} \right \|_{\ell_{1}/\ell_{2}}
\end{equation}
\begin{equation}
	TA-DistMult: f(s, p, o, t) = ( \bm{e}_{s} \circ \bm{e}_{o} ) \bm{e}_{p,t}^{\top},
\end{equation}
where $\circ$ is the element-wise product operation.

Inspired by diachronic word embeddings, Goel et al.~\cite{DeDistmult} proposed an alternative entity embedding function that dynamically projects entity-time pairs to hidden representations. For each entity and time pair $(s, t)$, the embedded entity vector $\bm{e}_{s, t}$ within the time period is computed as:
\begin{equation}
	\bm{e}_{s}[n]=\left\{\begin{array}{ll}
		\bm{a}_{\mathrm{s, t}}[n] \sigma\left(\bm{w}_{\mathrm{s}}[n] \mathrm{t}+\bm{b}_{\mathrm{s}}[n]\right), & \text { if } 1 \leq n \leq \gamma d \\
		\bm{a}_{\mathrm{s}}[n], & \text { if } \gamma d<n \leq d
	\end{array}\right.
\end{equation}
where $\bm{a}_{\mathrm{s}} \in \mathbb{R}^{d}$ and $\bm{w}_{\mathrm{s}}, \bm{b}_{\mathrm{s}} \in \mathbb{R}^{\gamma d}$ are learnable vectors associated with entity $s$, and $\sigma$ denotes an activation function. Using the above proposal, many existing traditional static KGE models, such as TransE, DistMult and SimplE, also constructed their temporal versions to deal with time information by replacing original $\bm{e}_{s}$ and $\bm{e}_{o}$ with $\bm{e}_{s, t}$ and $\bm{e}_{o, t}$, respectively.

TeRo~\cite{TeRo2020} defined the temporal evolution of entity embeddings as a rotation in complex vector space. For any timestamp $t$, TeRo transforms entities and relations to their corresponding complex embeddings, i.e., $\bm{e}_{s}$, $\bm{e}_{p}$, $\bm{e}_{o} \in \mathbb{C}^{k}$, and acquires time-specific entity embeddings $\bm{e}_{s, t}$ and $\bm{e}_{o, t}$ by regarding each timestamp $t$ as an element-wise rotation of the original time-independent entity embeddings. Thereafter, the relation embedding $\bm{e}_{p}$ is to consider as translation from the time-aware subject embedding $\bm{e}_{s, t}$ to the conjugate of the time-aware object embedding $\overline{\bm{e}}_{o, t}$. TeLM~\cite{TeLM2021} moved beyond complex-valued embeddings and utilized more expressive multivector representations from asymmetric geometric products to model entities, relations, and timestamps for temporal knowledge graph embedding.

Messner et al.~\cite{BoxTE2022} constructed a spatio-translational TKGE model based on the static box embedding method BoxTE~\cite{BoxE}. In BoxTE, each entity $e$ is translated to be associated with two vectors, a base position vector $\bm{e}_{s} \in \mathbb{R}^{d}$ and a translational bump vector $\bm{b}_{s} \in \mathbb{R}^{d}$. A corresponding time bump is then calculated by integrating relation $p$ and timestamp $t$ following the below formulation:
\begin{equation}
	\bm{e}_{p, t} = \bm{\alpha}_{p}\bm{K}_{t},
\end{equation}
where $\alpha$ indicates a scalar vector correlated to a relation and $\bm{K}_{t}$ is a time-dependent matrix. Finally, the head and tail entity representations are designed as:
\begin{equation}
	\begin{array}{l}
	\overline{\bm{e}}_{s} = \bm{e}_{s} + \bm{b}_{o} + \bm{e}_{p, t} \\
	\overline{\bm{e}}_{o} = \bm{e}_{o} + \bm{b}_{s} + \bm{e}_{p, t}.
	\end{array}
\end{equation}

In summary, many researchers have made great contributions in TKGE. However, their approaches only adopt random sampling strategies to generate negative facts by randomly selecting a candidate entity from the entity set $\mathbb{E}$ to replace the head or tail entity from the original positive triple. A more advanced sampling scheme can significantly improve the embedding performance. In this paper, we describe a new robust framework based on adversarial learning for improving the representation ability of TKGE models by constructing high quality plausible negative facts to train the discriminator. Compared with the above methods, our framework can generate more plausible negative samples, thereby improving the performance and practical usefulness of the temporal knowledge graph embedding model.

\section{The Proposed Framework}
\label{section3}
A temporal knowledge graph is a directed graph where nodes represent various entities, edges correspond to various relations between pairs of entities and each fact carries a time attribute which indicates the period during which this triple fact is valid. Given a temporal knowledge graph consisting of a collection of observed facts $(s, p, o, t)$, and a pre-defined embedding dimension $d$, temporal knowledge graph embedding projects each entity $s \in \mathbb{E}$ and relation $p \in \mathbb{R}$ into a $d$-dimensional continuous feature space. With this numerical vector representation, we can relieve the inherent issue of data sparsity and computational burden in large-scale TKG and support downstream applications such as link prediction and triple classification. 

\begin{figure*}[!htbp]
	\centering
	\includegraphics[width=12cm]{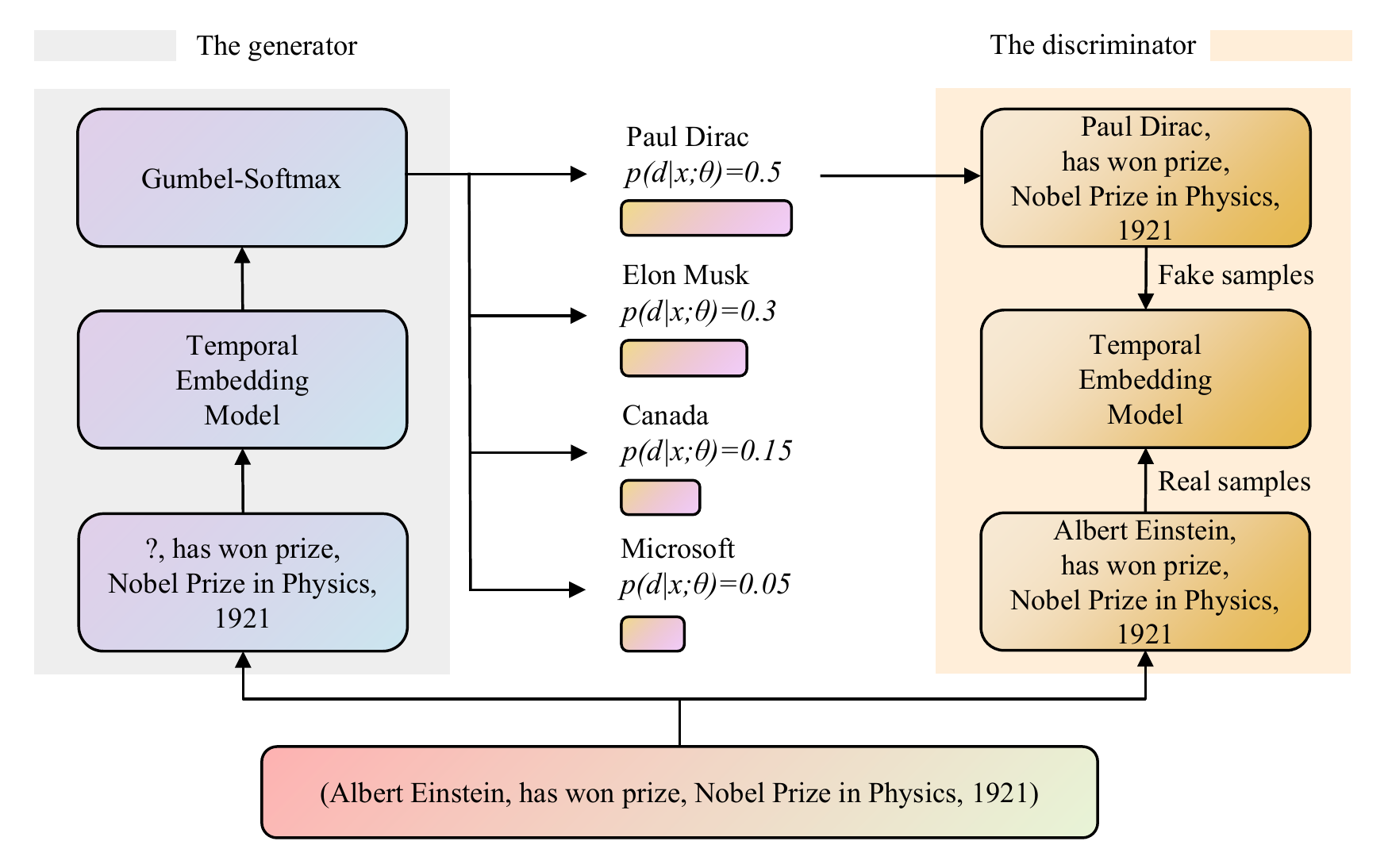}
	\caption{The framework of the proposed adversarial learning for temporal knowledge graph embedding. The generator learns to construct plausible negative facts for the discriminator. The discriminator utilizes the generated corrupted and valid triples to yield a robust and effective temporal knowledge graph representation.}
	\label{Figure2}
\end{figure*}

In this section, we elaborate our TKGE model in detail. Figure \ref{Figure2} delineates the proposed adversarial learning framework. At the beginning, a head or tail entity is discarded randomly over an authentic fact, and the resulting fragmentary fact $(?, has\ won\ prize, Nobel\ Prize\ in\ Physics, 1921)$ is obtained as the input of the generator. The generator receives it and selects another entity that has high similarity with $Albert\ Einstein$ (see $Paul\ Dirac$ in the figure) from a collection of candidate entities to construct a corrupted fact $(Paul\ Dirac, has\ won\ prize, Nobel\ Prize\ in\ Physics, 1921)$. For the discriminator direction, the generated and true facts are jointly fed into the TKGE model for learning robust representations. In consequence, we cannot only ensure that this framework can generate more diverse entities, but also guarantee that the generated entities are proximal in terms of semantics or functionality to the original ones in feature embedding space.

\subsection{Generator for sampling negative facts}
The goal of the generator is to construct more plausible negative facts for training the discriminator more effectively than what can be achieved through traditional random negative sampling methods.

\subsubsection{Deficiencies of traditional negative sampling}
Since Bordes et al.~\cite{transE2013} introduced uniform negative sampling to yield corrupted triple facts, many methods have followed this strategy to sample negative facts during model training. This strategy stochastically replaces the head or tail entity which form the original positive triple with a candidate entity from the entity set $\mathbb{E}$. All candidate entities in the entity set share the same probability of being chosen. Most TKGE models also follow this sampling pattern.

Unsurprisingly, this sampling approach is limited in its capacity of training an effective representation model in most cases. For instance, given a valid triple $(Albert\ Einstein, has\ won\ prize, Nobel\ Prize\ in\ Physics,\\ 1921)$, our purpose is to replace the head entity with another entity to constitute a negative triple. Considering the relation ``$ has\ won\ prize$'' and the entity type of ``$Albert\ Einstein$'', it is intuitive that the head entity should be a renowned individual. If we apply random sampling to select candidate entities, the constructed negative triple such as $(Microsoft, has\ won\ prize, Nobel\ Prize\ in\ Physics,\\ 1921)$ or $(Canada, has\ won\ prize, Nobel\ Prize\ in\ Physics,\\ 1921)$ can be trivially distinguished by the discriminator, resulting in infrequent parameter updates. In contrast, if we utilize a sampling scheme that generates negative samples with more reliability, such as $(Paul\ Dirac, has\ won\ prize, Nobel\ Prize\ in\ Physics, 1921)$, then it forces the discriminator to be more fully trained and further improve the representation ability of the embedding model.

To this end, we introduce a generative adversarial framework to refactor more plausible negative facts instead of traditional uniform random sampling strategy. Here, the generator is designed to construct reasonable negative triples, while the discriminator receives these high-quality training facts to further refine the embedding model. However, there is still a ``vanishing gradient'', also called ``non-differentiability'' issue in discrete data generation.

\subsubsection{Gumbel-Max reparametrization for discrete data}
We first establish the motivation why training generative adversarial learning model on discrete data is a pivotal problem from mathematical and instance perspectives. In the mathematical sense, assuming the total number of entities is $\left | E \right | $, the one-hot vector $\bm{y} \in \mathbb{R}^{\left | E \right |}$ which indexes the next candidate entity can be obtained via the generator by sampling:
\begin{equation}
	\bm{y} \sim \sigma(\log (\bm{\kappa})),
	\label{equation8}
\end{equation}
where $\log (\bm{\kappa}) \in \mathbb{R}^{\left | E \right |}$ indicates the output logits of the generator and $\sigma(\cdot)$ is the Softmax function. The sampling operation in Equation (\ref{equation8}) indicates a step function which is not differentiable at the last layer of the generator. Since the differential coefficient of a step function is 0 almost everywhere, we have ${\partial y} /  {\partial \theta_{G}}=0, \text { a.e. }$, where $\theta_{G}$ represents the parameters associated with the generator. Using the chain rule, the gradients of the generator $l_{G}$ with respect to $\theta_{G}$ are formulated as:
\begin{equation}
	\frac{\partial l_{G}}{\partial \theta_{G}}=\frac{\partial y}{\partial \theta_{G}} \frac{\partial l_{G}}{\partial y}=0 \quad \text { a.e. }
\end{equation}

Therefore, $\partial l_{G} / \partial \theta_{G}=0$ means that the gradient of the generator loss cannot be propagated back to the generator through the discriminator. In other words, the generator cannot update its own parameters based on the feedback mechanism provided by the discriminator. This circumstance is called the ``vanishing gradient'' or ``non-differentiability'' issue of generative adversarial networks in discrete data applications.

Let us consider a concrete example. Even though there are two disparate vectors, $\bm{\alpha} = [0.25, 0.35, 0.25, 0.15]$ and $\bm{\beta} = [0.05, 0.70, 0.15, 0.10]$, which are both obtained from the Softmax layer of the generator, the final one-hot results corresponding to these two vectors have not changed after the sampling operation, i.e., $one\_hot(\bm{\alpha}) = one\_hot(\bm{\beta}) = [0, 1, 0, 0]$. Therefore, identically sampled one-hot vectors are repetitively fed to the discriminator, making the gradients calculated by the discriminator ineffective, as the generator loses its convergence direction.

In order to prevent the vanishing gradient problem in discrete data, a reparametrization trick is introduced to separate the uncertainty of the discrete variables, making it possible to apply gradient propagation through intermediate nodes that were previously not differentiable. 

For a mathematical view, consider a computation graph consisting of a discrete sample variable $\bm{z}$ whose distribution depends on parameter $\theta$ and a loss function $f(\bm{z})$, the objective expected loss $L(\theta)=\mathbb{E}_{\bm{z} \sim p_{\theta}(\bm{z})}[f(\bm{z})]$ is anticipated to be minimized to achieve improved model performance via gradient propagation mechanisms, for which the estimate for $\nabla_{\theta} \mathbb{E}_{\bm{z} \sim p_{\theta}(\bm{z})}[f(\bm{z})]$ is prerequisite. However, there is no opportunity for end-to-end learning because of this discrete sampling step. In this scenario, we can leverage a reparametrization trick to deploy a deterministic function $g$ of the parameters $\theta$ and an independent random variable $\bm{\epsilon}$ to compute the discrete sample $\bm{z}$ directly, i.e., $\bm{z} = g(\theta, \bm{\epsilon})$. Hereafter, the path-wise gradients from $f$ to $\theta$ can be calculated smoothly without any obstacle:
\begin{equation}
\frac{\partial}{\partial \theta} \mathbb{E}_{\bm{z} \sim p_{\theta}}[f(\bm{z})]=\frac{\partial}{\partial \theta} \mathbb{E}_{\bm{\epsilon}}[f(g(\theta, \bm{\epsilon}))]=\mathbb{E}_{\bm{\epsilon} \sim p_{\bm{\epsilon}}}\left[\frac{\partial f}{\partial g} \frac{\partial g}{\partial \theta}\right].
\end{equation}

\begin{figure}
	\subfigure[$one\_hot$]{
		\begin{minipage}[htb]{0.25\linewidth}
			\centering
			\includegraphics[height=0.7in]{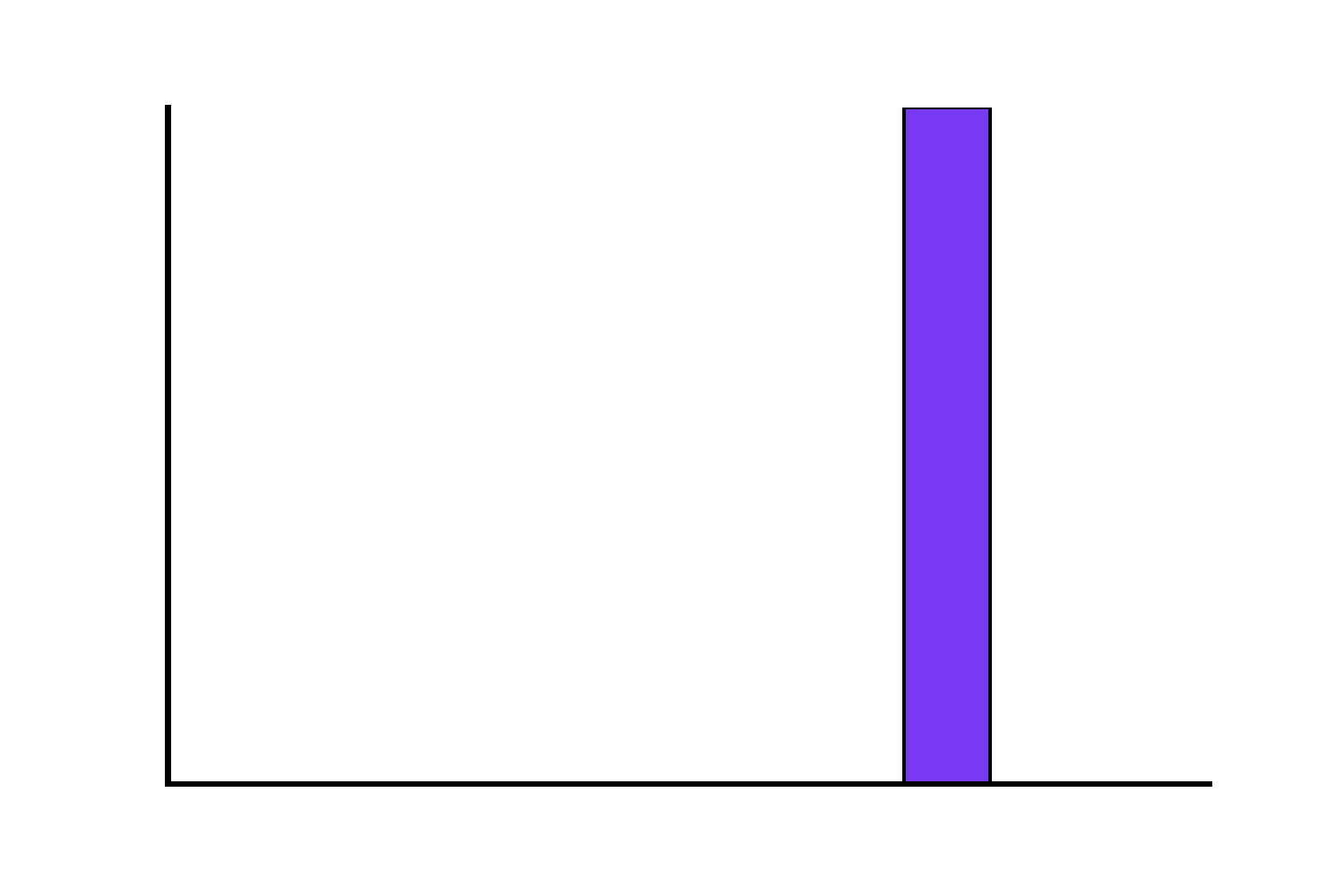}
		\end{minipage}
	}
	\subfigure[$\tau = 0.1$]{
		\begin{minipage}[htb]{0.25\linewidth}
			\centering
			\includegraphics[height=0.7in]{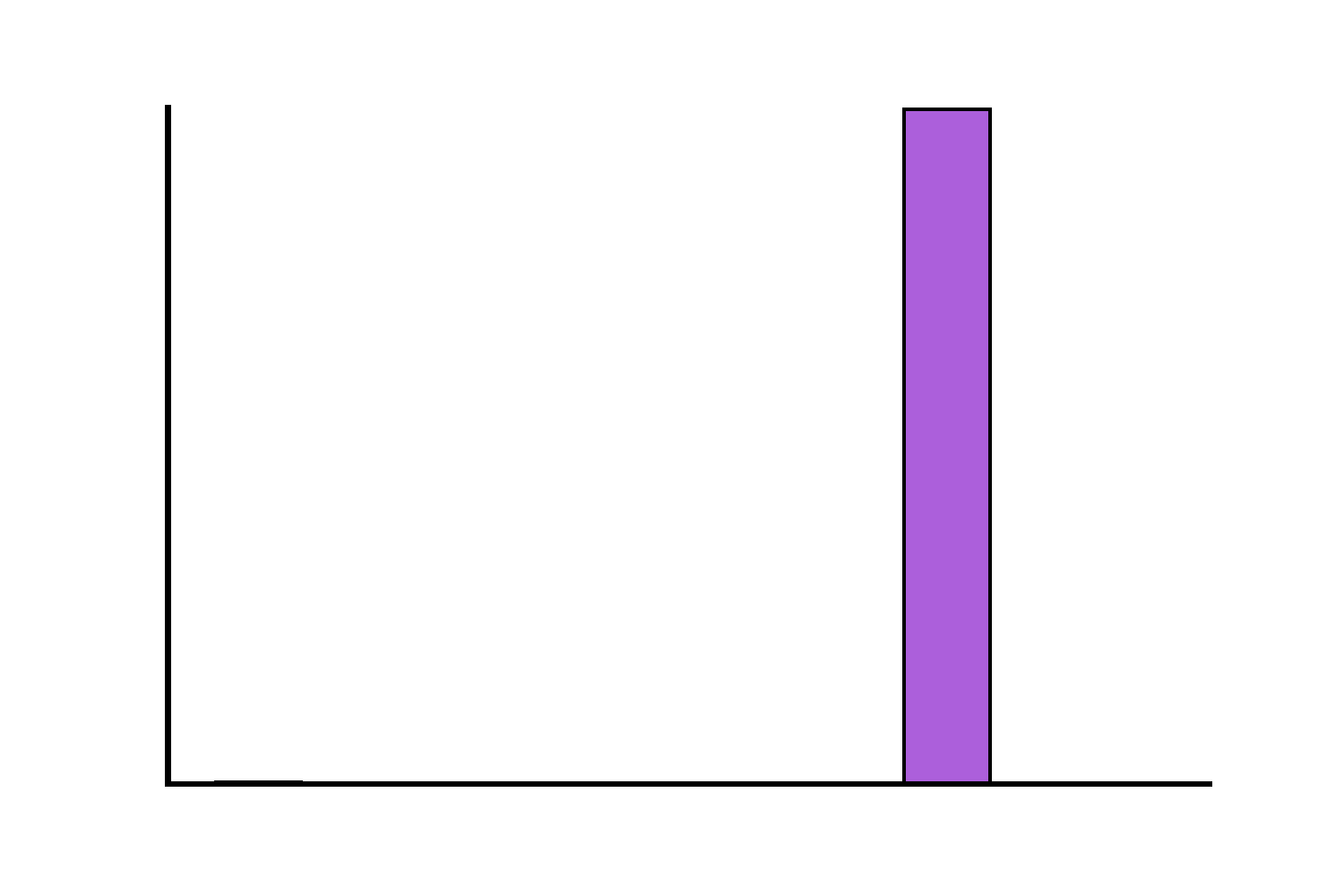}
		\end{minipage}
	}
	\subfigure[$\tau = 0.5$]{
		\begin{minipage}[htb]{0.25\linewidth}
			\centering
			\includegraphics[height=0.7in]{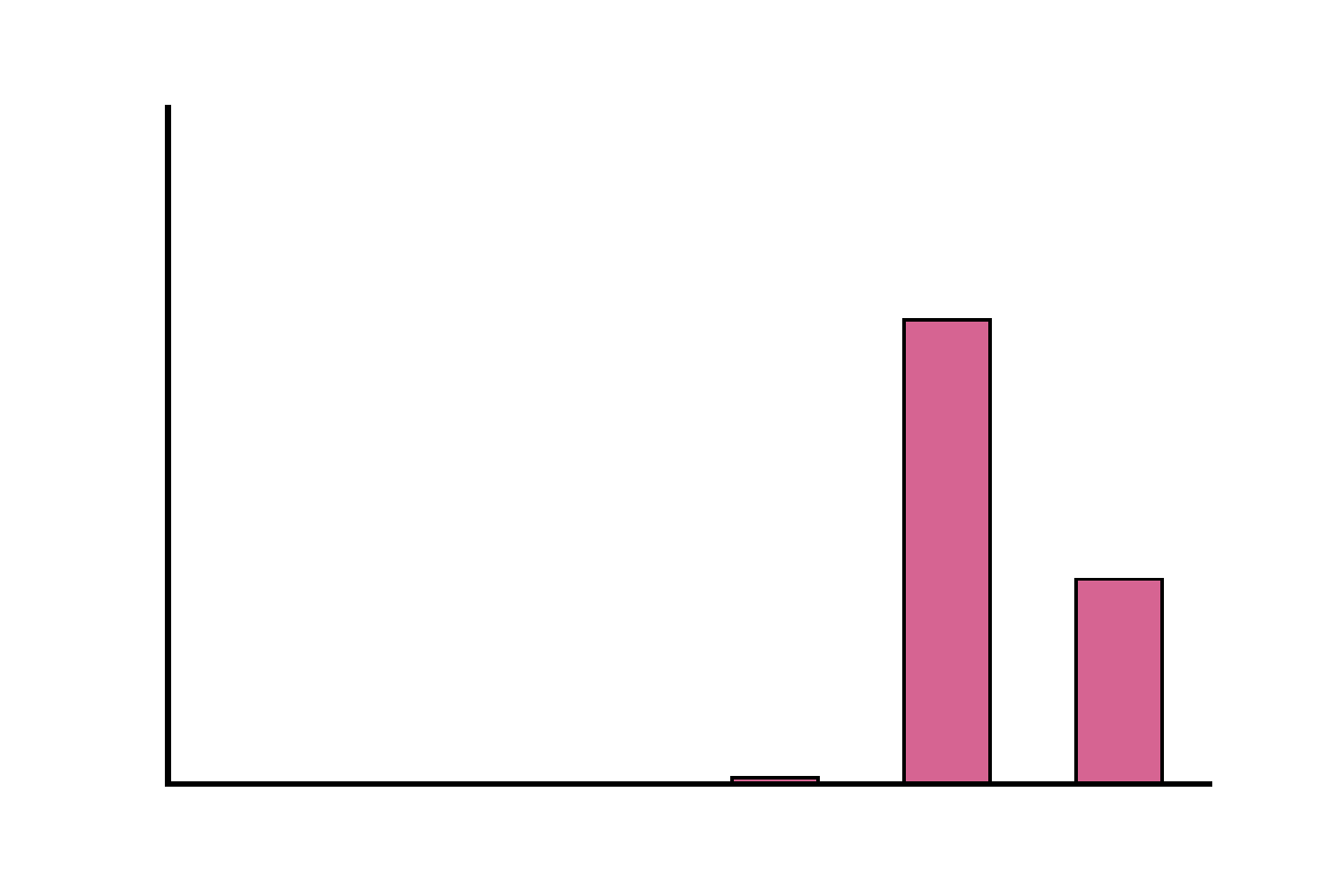}
		\end{minipage}
	}

	\subfigure[$\tau = 1$]{
		\begin{minipage}[htb]{0.25\linewidth}
			\centering
			\includegraphics[height=0.7in]{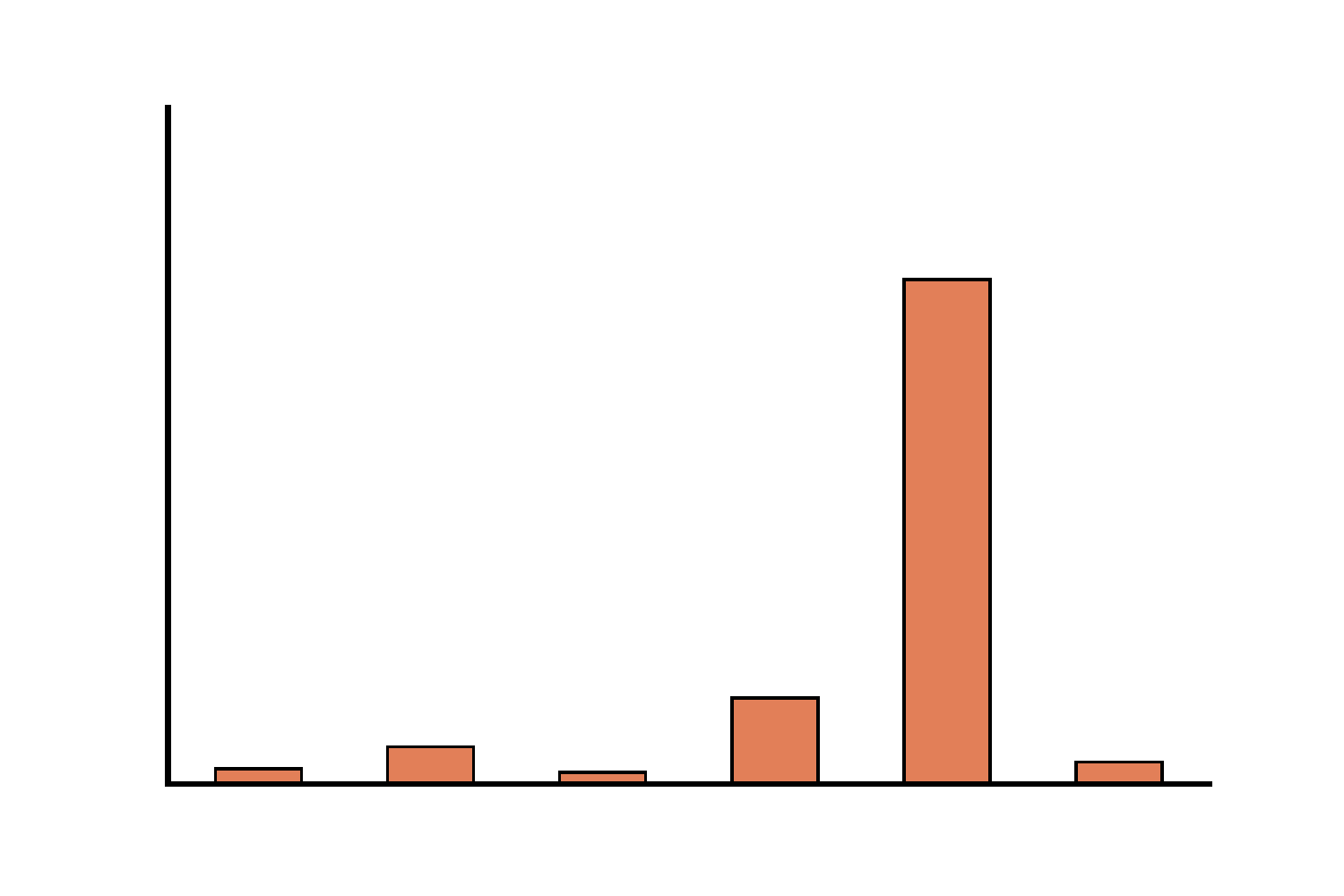}
		\end{minipage}
	}
	\subfigure[$\tau = 5$]{
		\begin{minipage}[htb]{0.25\linewidth}
			\centering
			\includegraphics[height=0.7in]{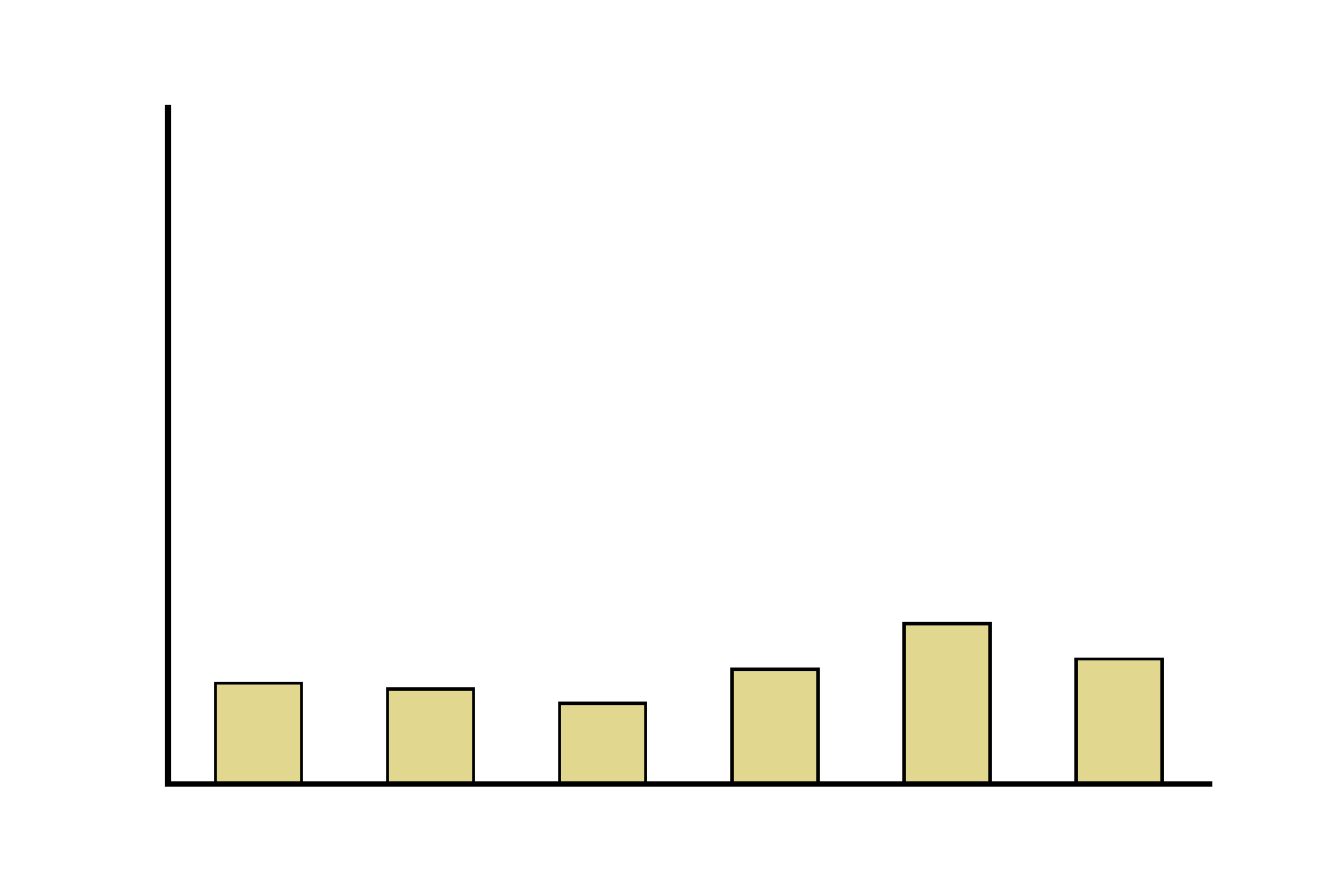}
		\end{minipage}
	}
	\subfigure[$\tau = 10$]{
		\begin{minipage}[htb]{0.25\linewidth}
			\centering
			\includegraphics[height=0.7in]{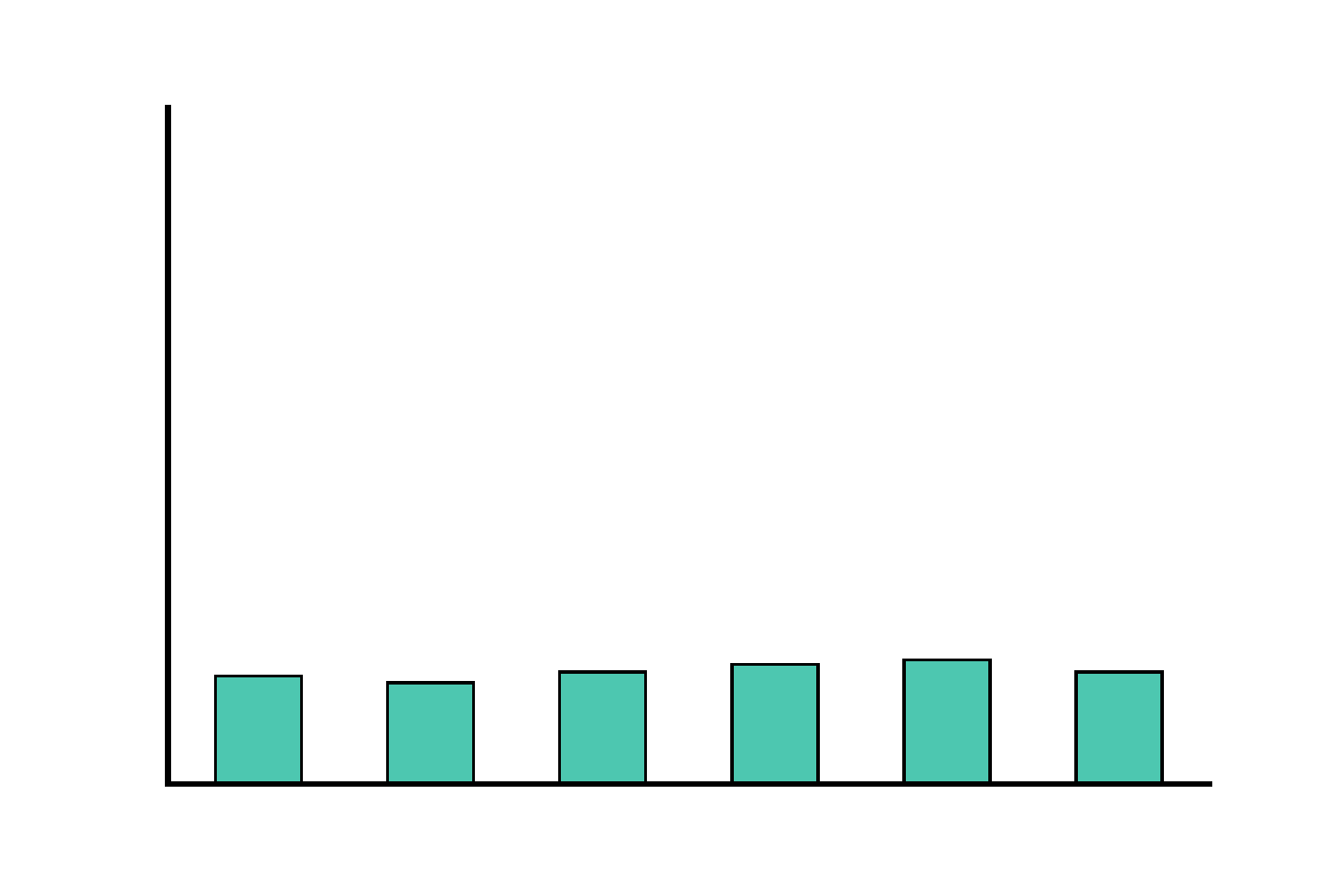}
		\end{minipage}
	}
	\caption{The influence of different values of temperature $\tau$ on the Gumbel-Softmax distribution. For low temperature, the expected value of a Gumbel-Softmax approaches the expected one\_hot encoding. As the temperature increases, the expected value converges to a uniform distribution.}
	\label{Figure3}
\end{figure}

As we mentioned before, discrete data has difficulty facilitating gradient propagation during the adversarial training procedure. Thus, the Gumbel-Max trick~\cite{jang2016categorical, maddison2016concrete}, a reparametrization technique, is employed to instead sample from a categorical distribution. Given the sampling probabilities for each category $\bm{\kappa}_{1}, \bm{\kappa}_{2}, ..., \bm{\kappa}_{\left | E \right |}$ which are as known as the output of the generator in our proposed framework, the samples can be expressed as:
\begin{equation}
	\bm{y}=one\_hot\left(\underset{1 \le i \le \left | E \right |}{\arg \max }\left[\log \bm{\kappa}_{i} + \bm{g}_{i} \right]\right),
	\label{equation 13}
\end{equation}
where $\bm{\kappa}_{i}$ is the $i$-th element of $\bm{\kappa}$ and $\bm{g}_{1}, \bm{g}_{2}, ..., \bm{g}_{\left | E \right |}$ indicate independent and identically distributed samples drawn from a standard Gumbel distribution, i.e., $\bm{g}_{i} \sim Gumbel(0,1)$. This distribution can be sampled utilizing inverse transform sampling by gaining $\bm{u} \sim Uniform(0, 1)$ and computing $\bm{g} = -\log (-\log(\bm{u}))$. So far the $\arg \max$ manipulation in Equation (\ref{equation 13}) is still non-differentiable. We relax the discreteness by applying the Softmax function as a continuous, differentiable approximation to further approximate $\arg \max$, and calculate a modified $\left | E \right |$-dimensional sample vector $\hat{\bm{y}}_{i}$:
\begin{equation}
	\hat{\bm{y}}_{i}=\frac{\exp \left(\left(\log \left(\bm{\kappa}_{i}\right)+\bm{g}_{i}\right) / \tau\right)}{\sum_{j=1}^{\left | E \right |} \exp \left(\left(\log \left(\bm{\kappa}_{j}\right)+\bm{g}_{j}\right) / \tau\right)}.
\end{equation}

Here, $\tau > 0$ is a controllable hyper-parameter referred to as the inverse temperature. As the temperature $\tau$ approaches $0$, samples from the Gumbel-Softmax distribution equal one-hot vectors and the Gumbel-Softmax distribution becomes identical to the categorical distribution. Figure \ref{Figure3} shows the influence of the temperature $\tau$ on sampling results. The probability density function of Gumbel-Softmax distribution is defined as follows:
\begin{equation}
	\small
	p_{\bm{\kappa}, \tau}\left(\hat{\bm{y}}_{1}, \ldots, \hat{\bm{y}}_{\left | E \right |}\right)=\Gamma(\left | E \right |) \tau^{\left | E \right |-1}\left(\sum_{i=1}^{\left | E \right |} \bm{\kappa}_{i} / \hat{\bm{y}}_{i}^{\tau}\right)^{-\left | E \right |} \prod_{i=1}^{\left | E \right |}\left(\bm{\kappa}_{i} / \hat{\bm{y}}_{i}^{\tau+1}\right).
\end{equation}

Now $\hat{\bm{y}}$ can be differentiated with respect to $\log (\bm{\kappa})$, we can directly apply this $\hat{\bm{y}}$ as the output of the generator, which also in turn serves as the input of the discriminator. The addition of the Gumbel-Softmax function ensures that the entire model can be continuously trained and improved using back propagation and chain rules. The generator can smoothly generate negative samples without worrying about the issue of vanishing gradients.

\subsection{TKG embedding discriminator}
The temporal knowledge graph embedding discriminators constructed in our framework are directly adopted from previous TKGE models, so that the performance differences of the same models on different training modes can be compared intuitively and clearly. As described in Section \ref{section2}, different temporal knowledge graphs have individual structures as well as scoring functions. Unlike previous embedding models where the negative samples were generated by random sampling from the entire set of entities, we apply an adversarial learning framework to construct more reasonable negative samples to refine the performance of embedding model. Compared with previous models, our framework does not need to modify the original embedding model's structure, but only changes the conventional sampling pattern via adversarial learning. In this fashion, the experimental results can be directly contrasted with the original model to verify the impact of the proposed framework on model performance.

\subsection{Generator architecture}
The focus of this article is to compare the impact of different negative sampling schemes on model performance under identical conditions. For the above objective, we utilize the existing TKGE methods directly in the discriminator, and also limit any modification to the embedding models in the generator to ensure that the experimental results solely reflect the effect of the sampling scheme.

Due to its representative and easily reproducible characters, TTransE is employed as the backbone of the generator. The input of this module is composed of four parts: an entity, a relation, a timestamp and a position indicator which is used to indicate whether the missing entity is a head or a tail entity. First, the entity, relation and timestamp are initially represented via specific feature space vectors through their respective embedding transformation matrix, the embedded vectors of entity, relation and timestamp are concatenated with the position indicator, and they are fed into a linear and Gumbel-Softmax layer to obtain the candidate entity. This entity is associated with the input of the generator including head entity and relation to form the corrupted fact. These generated negative samples are then fed into the discriminator along with positive samples for the final entity and relation embedding matrices.

\subsection{Training strategy}
The training procedure of the adversarial learning framework consists of two main aspects: (1) the update of parameters $\phi$ in the discriminator $D$; (2) the update of parameters $\theta$ in the generator $G$.

The discriminator network $D(\bm{x}; \phi)$ is designed to identify the authenticity of a given sample, i.e., to determine whether sample $\bm{x}$ is from the real distribution $p_r(\bm{x})$ or from the generated distribution $p_\eta(\bm{x})$. Using label $y = 1$ for positive samples and $y = 0$ for negative samples, the output of the discriminator represents the likelihood that the input $\bm{x}$ belongs to the true sample distribution. Given a sample $(\bm{x}, y)$, $y = \{0, 1\}$ denotes the authenticity of this sample. The goal of the discriminator network $D(\bm{x}; \phi)$ is minimizing the cross-entropy objective function:
\begin{equation}
	\begin{aligned}
		\min L_{D} = &\min _{\phi}- \left(\mathbb{E}_{\bm{x}}[y \log p(y=1 \mid \bm{x}) \right. \\ 
		& \left. + (1-y) \log p(y=0 \mid \bm{x})]\right).
	\end{aligned}
	\label{equa13}
\end{equation}

If the actual distribution $p(\bm{x})$ is a mixture of the real sample distribution $p_r(\bm{x})$ and the generated sample distribution $p_\eta(\bm{x})$ in equal proportion, then Equation (\ref{equa13}) can be reformulated as:
\begin{equation}
	\begin{aligned}
	\min L_{D}&=\min _{\phi}- (\mathbb{E}_{\bm{x} \sim p_{r}(\bm{x})}[\log D(\bm{x} ; \phi)]  \\
	& \ \quad +\mathbb{E}_{\bm{x}' \sim p_\eta(\bm{x}')}[\log (1-D(\bm{x}' ; \phi))]) \\
	& = \min_{\phi}- (\mathbb{E}_{\bm{x} \sim p_{r}(\bm{x})}[\log D(\bm{x} ; \phi)]\\
	&\ \quad +\mathbb{E}_{\bm{z} \sim p(\bm{z})}[\log (1-D(G(\bm{z}; \theta); \phi))]).
	\end{aligned}
\end{equation}
where $G(\bm{z} ; \theta)$ is the generator network. The objective of the generator network $G(\bm{x} ; \theta)$ is the opposite of the discriminator, which needs to make every effort to ``fool'' the discriminator into believing that the samples constructed by the generator are ``real'' facts. Therefore, its loss function is defined as:
\begin{equation}
	\begin{aligned}
		\min L_{G} &=\min_{\theta}\left(\mathbb{E}_{\bm{z} \sim p(\bm{z})}[\log (1-D(G(\bm{z} ; \theta) ; \phi))]\right) \\
		&= \max _{\theta}\left(\mathbb{E}_{\bm{z} \sim p(\bm{z})}[\log D(G(\bm{z} ; \theta) ; \phi)]\right).
	\end{aligned}
\end{equation}

Once we obtain the optimization objectives of the generator and the discriminator separately, we can merge them into a single whole. In this way, the objective function of the whole generative adversarial network can be viewed as a minimization maximization game,
\begin{equation}
	\begin{aligned}
	L_{G A N}=&\min_{\theta} \max _{\phi}\left(\mathbb{E}_{\bm{x} \sim p_{r}(\bm{x})}[\log D(\bm{x} ; \phi)] \right.  \\
	&\left. +\mathbb{E}_{\bm{x}' \sim p_{\eta}(\bm{x}')}[\log (1-D(\bm{x}' ; \phi))]\right)
	\end{aligned}
	\label{equa16}
\end{equation}

Compared with a single-objective optimization task, the generative adversarial structure fields two networks with opposing optimization goals. Seen together, practitioners need to spend a lot of effort to balance the performance of the two networks, which is why generative adversarial networks are brittle and difficult to train. In addition, the generative power should neither be too strong nor too weak for the generator. If it is too strong, it will make it difficult for the discriminator to distinguish truth from falsity, and if it is too weak, it will make it extremely easy for the discriminator to distinguish truth from falsity. Either situation leads to poor performance of the discriminator and may be further accompanied by pattern collapse, resulting in a lack of diversity in the generated samples.

The main reason for this phenomenon is that the original generative adversarial network uses Jensen-Shannon divergence (JS divergence) as a measure of similarity between the real and generated sample distributions. Assuming that the real and generated sample distributions are known, the optimal discriminator $ D^{\star}(\bm{x})$ can be formulated as:
\begin{equation}
	D^{\star}(\bm{x})=\frac{p_{r}(\bm{x})}{p_{r}(\bm{x})+p_{\eta}(\bm{x})}.
	\label{equa17}
\end{equation}

We can substitute the optimal discriminator $D^{\star}(\bm{x})$ of Equation (\ref{equa17}) into Equation (\ref{equa16}), and it becomes:
\begin{equation}
	\begin{aligned}
		L\left(G \mid D^{\star}\right) &=\mathbb{E}_{\bm{x} \sim p_{r}(\bm{x})}\left[\log D^{\star}(\bm{x})\right] \\
		&\quad +\mathbb{E}_{\bm{x} \sim p_{\eta}(\bm{x})}\left[\log \left(1-D^{\star}(\bm{x})\right)\right] \\
		&=\mathbb{E}_{\bm{x} \sim p_{r}(\bm{x})}\left[\log \frac{p_{r}(\bm{x})}{p_{r}(\bm{x})+p_{\eta}(\bm{x})}\right]\\
		&\quad+\mathbb{E}_{\bm{x} \sim p_{\eta}(\bm{x})}\left[\log \frac{p_{\eta}(\bm{x})}{p_{r}(\bm{x})+p_{\eta}(\bm{x})}\right] \\
		&=2 J S\left(p_{r}, p_{\eta}\right)-2 \log 2.
	\end{aligned}
\end{equation}

Here, $JS(\cdot)$ indicates the JS divergence. From the above formulation, it can be demonstrated that the generator loss defined in GANs can be transformed to minimize the JS divergence between the real and the generated sample distribution with the optimal discriminator. When the two distributions are the same, the JS divergence between them becomes $0$, so the corresponding loss is $L\left(G \mid D^{\star}\right)=-2 \log 2$. However, when the two distributions have no overlapping part or the overlapping parts are negligible, the JS divergence between them becomes fixed constant of $\log2$, and does not vary with the distance between the two distributions. This situation means the partial derivative (i.e., the gradient) of the loss function with respect to the generator parameters $\theta$ is $0$, i.e., $\partial L\left(G \mid D^{\star}\right) / \partial \theta=0$, causing the gradient to vanish. This issue makes the original GAN difficult to train, the generator model collapses, and the generated negative samples lack diversity.

The above section argues that the JS divergence is not an appropriate choice for measuring the distance between the real sample distribution $p_r(\bm{x})$ and the generated sample distribution $p_\eta(\bm{x})$. Inspired by Wasserstein GANs~\cite{gulrajani2017improved}, this paper utilizes Wasserstein distance (also known as Earth Mover distance) as a more robust and efficient measure to replace the original JS divergence, and thus optimizes the whole adversarial network. Given a real sample distribution $p_r(\bm{x})$ and a generated sample distribution $p_\eta(\bm{x})$, the Wasserstein distance between them can be defined as:
\begin{equation}
	W\left(p_{r}, p_{\eta}\right)=\inf _{\gamma \sim \Pi\left(P_{r}, P_{\eta}\right)} \mathbb{E}_{(\bm{x}, \bm{y}) \sim \gamma}[|\bm{x}-\bm{y}|],
	\label{equa19}
\end{equation}
where $\Pi\left(P_{r}, P_{\eta}\right)$ is the set of all possible joint distributions with marginal distribution $\gamma(\bm{x}, \bm{y})$. The difference between the Wasserstein distance and the JS divergence is that the JS divergence is constant when there is no or only negligible overlap between the two distributions, but the Wasserstein distance has the capability of varying with the distance between the two distributions without overlapping, making it more suitable for our method.

Although Wasserstein distance has so many advantages, Equation (\ref{equa19}) cannot be calculated directly, it needs to be converted into a solvable form by the Kantorovich-Rubinstein duality theorem~\cite{villani2008optimal}. According to this theorem, the Wasserstein distance can be transformed into an upper bound on the expected difference between these two distributions for a function, satisfying the K-Lipschitz continuum. Equation (\ref{equa19}) can be rewritten as:
\begin{equation}
	\bm{W}\left(p_{r}, p_{\eta}\right)=\frac{1}{K} \sup _{\|f\|_{L} \leq K}\left(\mathbb{E}_{x \sim p_{r}(\bm{x})}[f(\bm{x})]-\mathbb{E}_{x \sim p_{\eta}(\bm{x})}[f(\bm{x})]\right),
	\label{equation23}
\end{equation}
where $f(\cdot)$ is the K-Lipschitz function, that satisfies the following condition:
\begin{equation}
	\|f\|_{L} \triangleq \sup _{x \neq y} \frac{|f(\bm{x})-f(\bm{y})|}{|\bm{x}-\bm{y}|} \leq K.
\end{equation}

A function is a Lipschitz continuous function if it is differentiable and its derivatives are bounded. Because the discriminator network $D(\bm{x} ; \phi)$ (the expected TKGE model) satisfies the above conditions, it enables us to approximate the upper bound in Equation (\ref{equation23}) to :
\begin{equation}
	\begin{aligned}
		\min L_{D} &= \min _{\phi}-\left(\mathbb{E}_{\bm{x} \sim p_{r}(\bm{x})}[D(\bm{x} ; \phi)]-\mathbb{E}_{\bm{x} \sim p_{\eta}(\bm{x})}[D(\bm{x} ; \phi)]\right) \\
		& = \min _{\phi}-\left(\mathbb{E}_{\bm{x} \sim p_{r}(\bm{x})}[D(\bm{x} ; \phi)]-\mathbb{E}_{\bm{z} \sim p(\bm{z})}[D(G(\bm{z} ; \theta) ; \phi)]\right).
	\end{aligned}
\end{equation}

The goal of the generator network $L_{G}$ is to minimize the Wasserstein distance between the real distribution $p_{r}(\bm{x})$ and the generated distribution $p_{\eta}(\bm{x})$, letting the constructed negative samples achieve the highest possible discriminator scores. The optimization objective of the generator is shown below:
\begin{equation}
\min L_{G} =  \min _{\theta}-\mathbb{E}_{\bm{z} \sim p(\bm{z})}[D(G(\bm{z} ; \theta) ; \phi)].
\end{equation}

\begin{algorithm}
	\caption{The framework based on adversarial training with Wasserstein distance for temporal knowledge graph embeddings.}
	\label{al1}
	\SetKwInOut{Input}{Input}\SetKwInOut{Output}{Output}
	
	\Input{The set of positive facts $\Omega = \{(s, p, o, t)\}$, the number of training iterations $e$, the number of discriminator iterations per generator iteration $n_{dis}$, mini-batch size $m$, the learning rate of the generator $\alpha$, the learning rate of the discriminator $\beta$.}
	\Output{Entities and relations embeddings learned by the discriminator $D$.}
	\BlankLine
	\textbf{Initialize} the generator $G$ with parameters $\theta_{0}$, the discriminator $D$ with parameters $\phi_{0}$\;
	\For{$k = 1, \cdots, e$}{
		\For{$n = 1, \cdots, n_{dis}$}{
			\For{$i = 1, \cdots, m$}{
				$L^{(i)}_{D} = -[D(\bm{x}^{(i)};\phi)-D(G(\bm{z}^{(i)}; \theta) ; \phi) ]$ \tcp*[l]{Update the discriminator by minimizing $L^{(i)}_{D}$.}
			}
			$f_{\phi} \leftarrow \nabla_{\phi} \frac{1}{m} \sum_{i=1}^{m} L^{(i)}_{D}$\;
			$\phi \leftarrow \phi - \beta \cdot \operatorname{Adagrad}\left( \phi, f_{\phi} \right)$\;}
		\For{$i = 1, \cdots, m$}{
			$L^{(i)}_{G} = -D(G(\bm{z}^{(i)} ; \theta) ; \phi)$\tcp*[l]{Update the generator by minimizing $L^{(i)}_{G}$.}
		}
		$h_{\theta} \leftarrow \nabla_{\theta} \frac{1}{m} \sum_{i=1}^{m} L^{(i)}_{G}$\;
		$\theta \leftarrow \theta - \alpha \cdot \operatorname{Adagrad}\left( \theta, h_{\theta} \right)$\;
	}
	\textbf{Return} Entity and relation embeddings incorporating temporal information.
\end{algorithm}

Because $D(\bm{x} ; \phi)$ is an unsaturated function, the gradient of the generator network parameter $\theta$ does not vanish, which theoretically solves the problem of unstable training of the original GAN and alleviates the mode collapse problem to a certain extent, making the generated samples more plausible and diverse. The detailed training process of this adversarial framework for temporal knowledge graph embedding is described in Algorithm \ref{al1}.

\section{Experiments and analysis}
\label{section4}
In this section, we first introduce five experimental datasets in detail, then describe important parameter settings and basic comparison methods for our experiment. Afterwards, a link prediction task is constructed to compare and verify performance of the presented framework with benchmarks and state-of-the-art models. Next, the presentation of experimental results and corresponding quantitative analysis prove the validity of our model. Finally, we provide a visual inspection of qualitative example outputs to allow for a more intuitive comprehension of the capacities of our framework.
\begin{table*}[htb]
	\centering
	\caption{Statistics of the data sources.}
	\renewcommand{\arraystretch}{1.3}
	\begin{tabular}{l|cccccccc}
		\hline
		& \#Entities & \#Relations & \#Timestamps & \#Train & \#Valid & \#Test & Peroid & Instance of timestamps \\ \hline
		ICEWS14     & 7,128  & 230 & 365   & 72,826    & 8,963   & 8,941   & 1 year     & 2014-01-11   \\
		ICEWS05-15  & 10,488 & 251 & 4,017 & 386,962   & 46,092  & 46,275  & 11 years   & 2012-05-19   \\
		GDELT       & 500    & 20  & 366   & 2,735,685 & 341,961 & 341,961 & 1 year     & 2015-10-08   \\ \hline
		Yago11k     & 10,623 & 10  & 118   & 16,406    & 2,050   & 2,051   & 3,275 years & (1784, 1790) \\
		Wikidata12k & 12,554 & 24  & 125   & 32,497    & 4,062   & 4,062   & 2,001 years & (1826, 1840) \\ \hline
	\end{tabular}
	\label{table1}
\end{table*}

\subsection{Datasets}
We evaluate our framework on five large temporal knowledge graph embedding benchmark datasets, namely ICEWS14, ICEWS05-15, GDELT, Yago11k and Wikidata12k. Table \ref{table1} gives a summary of key dataset statistics. A detailed description of these datasets is shown below:
\begin{itemize}
	\item \textbf{ICEWS14} and \textbf{ICEWS05-15}~\cite{TAdistmult} are the two most common TKG benchmarks extracted from the large-scale event-based database, \textit{Integrated Crisis Early Warning System} (ICEWS), which curates socio-political temporal facts starting from 1995. ICEWS14 contains political events with specific time points in 2014, and ICEWS05-15 stores occurrences during 2005-2015. These two datasets are constructed by seleceting the most frequently occurring entities in the graph.
	
	\item \textbf{GDELT}~\cite{Know-evolve} is derived from a larger \textit{Global Database of Events,	Language, and Tone} (GDELT) knowledge graph that contains temporal facts about human behavior starting from 1979. This benchmark dataset is a subset of the unabridged GDELT. It contains facts with time annotations between April 1, 2015 and March 31, 2016, and only absorbs quadruple facts involving the 500 most frequent entities and 20 most common relations.
	
	\item \textbf{Yago11k} and \textbf{Wikidata12k}~\cite{hyte} are Yago3 and Wikidata subsets with temporal information filtered from large general knowledge graphs. The events recorded in Yago11k span a period of $3,275$ years, the period in Wikidata12k is $2,001$ years. 
\end{itemize}

As shown in Table \ref{table1}, the time span of ICEWS14, ICEWS05-15 and GDELT is 1 year or 11 years, we can apply one day as the time granularity to obtain fine-grained timestamps for this category of datasets with a short period. However, different from these datasets, the time span of Yago11k and Wikidata12k is much greater. Furthermore, there is an imbalance issue that might occur in terms of the number of facts in a particular interval. For example, relatively few events are recorded in Wikidata12k's, $[1791-1815]$ period, while some intervals (for example the single year 2012) are populated with numerous facts. Uniform division of timestamps by day or year would lead to considerable dispersion of timestamps, which is not conducive to effective model training. 

We follow the time-division strategy utilized in TeLM~\cite{TeLM2021} that defines a minimum threshold to guarantee there will be enough events in each time interval for Yago11k and Wikidata12k. As illustrated in Table \ref{table1}, we finally divide timestamps in 118 buckets in Yago11k, and 125 buckets in Wikidata12k.

\subsection{Comparison methods}
In order to comprehensively evaluate whether our proposed framework is effectively and consistently improving the performance of existing models, we select several representative TKGC models as baselines to be compared with our framework. These benchmarks are described as follows:
\begin{itemize}
	\item \textbf{TTransE}~\cite{ttranse} represents entities, relations and timestamps in a uniform low-dimensional feature space, and regards relations as translation calculations to concatenate the entities and timestamps.
	
	\item \textbf{TA-DistMult}~\cite{TAdistmult} leverages LSTM networks to transform temporal information to time-aware representations, then the embedding vectors process relations to form time-aware relation embeddings for score function.
	
	\item \textbf{De-SimplE}~\cite{DeDistmult} proposes an alternative temporal entity embedding model in which each fact and its associate timestamp are dynamically projected to time-aware hidden vector.
	
	\item \textbf{TeRo}~\cite{TeRo2020} defines the time-aware evolution of entity embeddings as a rotation in complex vector space, it acquires time-specific entity embeddings by interpreting timestamps as an element-wise rotation.
	
	\item \textbf{TeLM}~\cite{TeLM2021} employs multivector embeddings from asymmetric geometric products to model entities, relations, and timestamps for temporal knowledge graph representation.
	
	\item \textbf{BoxTE}~\cite{BoxTE2022} integrates relations and timestamps via a time-dependent matrix, and follows static box embedding patterns to accomplish the representation of temporal knowledge graph.

\end{itemize}

\subsection{Evaluation Metrics}
We evaluate our proposed framework on a link prediction task over the above-described temporal knowledge graph benchmarks. The goal of this characteristic task is to infer a held-out entity when given an existing entity, relation and timestamp query. More precisely, the target of time-aware link prediction is to predict the missing head entity $s$ if given $(p, o, t)$ or predict tail entity $o$ given $(s, p, t)$. Results are acquired by ranking discriminator scores.

For each quadruple $(s, p , o, t)$ in the test dataset, the true head entity (or tail entity) is circularly replaced by all entities in the entity set $\mathbb{E}$. Then, the scores associated with all quadruples are calculated, all scores are ranked in descending order. However, some reconstructed quadruples might coincidentally be authentic in original TKG, resulting in an incorrect assessment. In order to avoid this situation, following the convention used in most previous related studies, we apply the ‘Filtered’ setting to eliminate all reconstructed quadruples which incidentally exist either in the training, validation, or test datasets. Finally, there are three major criteria to measure model performance, low \textbf{MR}, high \textbf{MRR} and high \textbf{Hits@N\%} scores indicate good performance:
\begin{itemize}
	\item \textbf{MR}: the average rank of the real entities.
	
	\item \textbf{MRR}: the mean reciprocal rank of the real entities.
	
	\item \textbf{Hits@N\%}: the proportion of authentic entities that ranked in top $N$. Here, we particularly report the $N = 1, 3, 10$ scores to validate the performance of compared methods.
\end{itemize}

\subsection{Training protocol}
We implement baseline TKGE models and our proposed framework in PyTorch, and utilize a self-adaptive Adagrad optimizer for the training stage. It is worth noting that the primary purpose of the experiment is to demonstrate that our framework has the capability of effectively improving the performance of the benchmark TKGE models with strong generalization. Therefore, in this experiment, we do not set out to achieve optimal performance of each single benchmark model via exhaustive hyper-parameter tuning, but rather ensure that hyperparameters take the same value in different benchmark models as much as possible. Under such a standardized architecture, the performance of the models can be more clearly and intuitively represented.

In our experiments, we choose the well known TTransE model as the backbone of the generator in the adversarial framework, and utilize the various above-mentioned embedding methods as the discriminator directly. The main hyper-parameters in this experiment include the learning rate of the generator $\alpha$, the learning rate of the discriminator $\beta$, the dimension of embedding vectors $d$, the mini-batch size $m$, the number of overall training iterations $e$ and the number of discriminator training iterations per generator iteration $n_{dis}$. Following a similar setting in previous research~\cite{DeDistmult}, on ICEWS14, ICEWS05-15, Yago11k and Wikidata12k datasets, the parameter configurations are $\{\alpha = 0.001, \beta = 0.0001, d = 100, m = 512, e = 500, n_{dis} = 5\}$. On GDELT, the parameter configurations are $\{\alpha = 0.01, \beta = 0.001, d = 100, m = 512, e = 500, n_{dis} = 5\}$. We select the model via validating every 100 epochs which gives the best validation MRR.

\subsection{Evaluation results}

\begin{table*}[htbp]
	\centering
	\caption{Evaluation results on ICEWS14, ICEWS05-15 and GDELT datasets. The best two results among all methods are highlighted in {\color{red}red} and {\color{blue}blue}, and are formatted in boldface. 'F-' indicates those models that were trained via our proposed framework.}
	\renewcommand{\arraystretch}{1.3}
	\setlength\tabcolsep{1.8pt}
	\begin{tabular}{l|c@{\hspace{5pt}}cccc|c@{\hspace{5pt}}cccc|c@{\hspace{5pt}}cccc}
		\hline
		\multicolumn{1}{c|}{} & \multicolumn{5}{c}{ICEWS14}                  & \multicolumn{5}{c}{ICEWS05-15}               & \multicolumn{5}{c}{GDELT}                   \\ \cline{2-16} 
		\multicolumn{1}{c|}{\multirow{-2}{*}{Models}} &
		MR &
		MRR &
		Hits@1 &
		Hits@3 &
		Hits@10 &
		MR &
		MRR &
		Hits@1 &
		Hits@3 &
		Hits@10 &
		MR &
		MRR &
		Hits@1 &
		Hits@3 &
		\multicolumn{1}{c}{Hits@10} \\ \hline
	TTransE~\cite{ttranse}              & 186.58 & 0.2322 & 0.0502 & 0.3356 & 0.5807 & 146.08 & 0.2463 & 0.0598 & 0.3585 & 0.5919 & 94.00 & 0.0879 & 0.0130 & 0.0976 & 0.2281 \\
	TA\_DistMult~\cite{TAdistmult}          & 178.05 & 0.2938 & 0.1408 & 0.3705 & 0.5967 & 132.54 & 0.3474 & 0.1957 & 0.4180 & 0.6576 & 70.56 & 0.1676 & 0.0886 & 0.1773 & 0.3203 \\
	DE\_SimplE~\cite{DeDistmult}            & 272.91 & \color{blue}{\textbf{0.4387}} & \color{blue}{\textbf{0.3214}} & 0.4955 & 0.6672 & 150.53 & 0.4513 & 0.3301 & 0.5083 & 0.6949 & 65.03 & 0.1967 & 0.1165 & 0.2088 & 0.3592 \\
	TeRo~\cite{TeRo2020}                  & 177.21 & 0.4091 & 0.2848 & 0.4689 & 0.6572 & 119.71 & 0.4276 & 0.3010 & 0.4900 & 0.6789 & 68.01 & 0.1710 & 0.0889 & 0.1875 & 0.3281 \\
	TeLM~\cite{TeLM2021}                  & 213.12 & 0.4125 & 0.3033 & 0.4886 & 0.6625 & 137.10 & 0.4626 & 0.3433 & \color{blue}{\textbf{0.5287}} & \color{blue}{\textbf{0.7126}} & 65.26 & 0.1933 & 0.1188 & 0.2109 & 0.3535 \\
	BoxTE~\cite{BoxTE2022}                 & 239.36 & 0.4020 & 0.2947 & 0.4798 & 0.6599 & 124.83 & 0.4328 & 0.3146 & 0.4910 & 0.6846 & 66.83 & \color{blue}{\textbf{0.2295}} & \color{blue}{\textbf{0.1353}} & \color{blue}{\textbf{0.2359}} & \color{blue}{\textbf{0.3812}} \\ \hline
	F-TTransE             & \color{red}{\textbf{167.64}} & 0.2381 & 0.0481 & 0.3487 & 0.5966 & 144.73 & 0.2483 & 0.0567 & 0.3617 & 0.5957 & 91.36 & 0.1068 & 0.0199 & 0.1095 & 0.2401 \\
	F-TA\_DistMult        & \color{blue}{\textbf{173.20}} & 0.3476 & 0.2161 & 0.3992 & 0.6259 & 113.74 & 0.4011 & 0.2632 & 0.4685 & 0.6713 & 69.09 & 0.1727 & 0.0949 & 0.1977 & 0.3322 \\
	F-DE\_SimplE          & 296.79 & \color{red}{\textbf{0.4504}} & \color{red}{\textbf{0.3237}} & \color{red}{\textbf{0.4976}} & \color{red}{\textbf{0.6753}} & 182.06 & \color{blue}{\textbf{0.4680}} & \color{blue}{\textbf{0.3475}} & 0.5241 & 0.6990 & \color{blue}{\textbf{62.43}} & 0.2141 & 0.1280 & 0.2197 & 0.3789 \\
	F-TeRo                & 176.85 & 0.4222 & 0.3075 & 0.4836 & 0.6694 & \color{blue}{\textbf{112.23}} & 0.4392 & 0.3154 & 0.5068 & 0.6976 & 65.28 & 0.1859 & 0.0958 & 0.1978 & 0.3310 \\
	F-TeLM                & 185.01 & 0.4323 & 0.3022 & \color{blue}{\textbf{0.4956}} & \color{blue}{\textbf{0.6721}} & 126.58  & \color{red}{\textbf{0.4768}} & \color{red}{\textbf{0.3568}} & \color{red}{\textbf{0.5378}} & \color{red}{\textbf{0.7264}} & \color{red}{\textbf{61.11}} & 0.2024 & 0.1273 & 0.2185 & 0.3625 \\
	F-BoxTE               & 192.10 & 0.4277 & 0.2986 & 0.4817 & 0.6642 & \color{red}{\textbf{108.24}} & 0.4424 & 0.3166 & 0.5091 & 0.6887 & 62.77 & \color{red}{\textbf{0.2420}} & \color{red}{\textbf{0.1473}} & \color{red}{\textbf{0.2476}} & \color{red}{\textbf{0.4018}} \\ \hline
\end{tabular}
\label{table2}
\end{table*}

\begin{table*}[hbtp]
	\centering
	\caption{Evaluation results on Yago11k and Wikidata12k datasets. The best two results among all methods are highlighted in {\color{red}red} and {\color{blue}blue}, and are formatted in boldface. 'F-' indicates those models that were trained via our proposed framework.}
	\renewcommand{\arraystretch}{1.3}
	\begin{tabular}{l|ccccc|ccccc}
		\hline
		\multicolumn{1}{c|}{} & \multicolumn{5}{c|}{Yago11k}                & \multicolumn{5}{c}{Wikidata12k}            \\ \cline{2-11} 
		\multicolumn{1}{c|}{\multirow{-2}{*}{Models}} &
		MR &
		MRR &
		Hits@1 &
		Hits@3 &
		Hits@10 &
		MR &
		MRR &
		Hits@1 &
		Hits@3&
		Hits@10 \\ \hline
		TTransE~\cite{ttranse}               & \color{blue}{\textbf{796.16}}  & 0.1047 & 0.0299 & 0.1265 & 0.2395 & 490.11 & 0.2308 & 0.1340  & 0.2454 & 0.4363 \\
		TA\_DistMult~\cite{TAdistmult}          & 867.27  & 0.1260 & 0.0900   & 0.1350  & 0.2278 & 308.07 & 0.2294 & 0.1321  & 0.2501 & 0.4422 \\
		DE\_SimplE~\cite{DeDistmult}            & 1651.27 & 0.1401 & 0.0912 & 0.1388 & 0.2429 & 483.68 & 0.2370  & 0.1418 & 0.2554 & 0.4446 \\
		TeRo~\cite{TeRo2020}                  & 970.76  & 0.1320 & 0.0854 & 0.1371  & 0.2259 & 401.18 & 0.2105 & 0.1459 & 0.2465 & 0.4372 \\
		TeLM~\cite{TeLM2021}                  & 948.50  & 0.1433 & 0.0856 & 0.1413 & 0.2363 & 357.36 & 0.2420  & \color{blue}{\textbf{0.1552}} & 0.2665 & 0.4610  \\
		BoxTE~\cite{BoxTE2022}                 & 1107.44 & 0.1334 & 0.0831 & 0.1347 & 0.2330  & 467.50  & 0.2231 & 0.1429 & 0.2550  & 0.4438 \\ \hline
		F-TTransE             & \color{red}{\textbf{678.86}}  & 0.1115 & 0.0357 & 0.1353 & \color{blue}{\textbf{0.2465}} & \color{blue}{\textbf{184.01}} & 0.2433 & 0.1402 & 0.2651 & 0.4502 \\
		F-TA\_DistMult        & 1207.56 & 0.1378 & 0.0886 & 0.1401 & 0.2324 & \color{red}{\textbf{181.07}} & 0.2378 & 0.1436 & 0.2566 & 0.4561 \\
		F-DE\_SimplE          & 1632.77 & \color{red}{\textbf{0.1495}} & \color{blue}{\textbf{0.0969}} & 0.1421 & \color{red}{\textbf{0.2493}} & 503.45 & \color{blue}{\textbf{0.2474}} & 0.1414 & \color{blue}{\textbf{0.2737}} & 0.4631 \\
		F-TeRo                & 1032.86 & 0.1407 & 0.0904 & \color{blue}{\textbf{0.1466}} & 0.2402 & 410.08 & 0.2345 & 0.1477 & 0.2597 & 0.4428 \\
		F-TeLM                & 859.74  & \color{blue}{\textbf{0.1481}} & \color{red}{\textbf{0.0999}} & \color{red}{\textbf{0.1511}} & 0.2450  & 320.73 & \color{red}{\textbf{0.2535}} & \color{red}{\textbf{0.1608}} & \color{red}{\textbf{0.2817}} & \color{red}{\textbf{0.4808}} \\
		F-BoxTE               & 1044.65 & 0.1395 & 0.0886 & 0.1439 & 0.2387 & 434.96 & 0.2319 & 0.1442 & 0.2659 & \color{blue}{\textbf{0.4682}} \\ \hline
	\end{tabular}
\label{table3}
\end{table*}

Table \ref{table2} gives a detailed comparison of the proposed framework and comparative methods on ICEWS14, ICEWS05-15 and GDELT datasets. We can observe that:
\begin{itemize}
	\item On these three datasets, the temporal knowledge graph embedding approaches trained via our adversarial framework (indicated as 'F-') gain a significant performance boost on all evaluation metrics compared with the corresponding original methods.
	
	\item F-DE\_SimplE and F-TeLM show top performance on ICEWS14 and ICEWS05-15, F-BoxTE outperforms other methods on GDELT. Indeed, as shown in Table \ref{table1}, GDELT is substantially larger than both ICEWS datasets. Such results also illustrate that BoxTE can capture more temporal patterns and conducive information on large-scale datasets.
	
	\item The early TKG representation learning models, TTransE and TA\_DistMult perform poorly in terms of performance compared with current methods, due to their inherent limitations in expressiveness. However, the insights from their works have inspired later researchers to achieve continuous improvements.
	
	\item On ICEWS datasets, the proposed framework can improve the performance by an average of 2.3 and 2 points of MRR beyond the original methods on ICEWS14 and ICEWS05-15, respectively. Even under the increased scale of the GDELT that includes 2.7 million training facts, there is still an average improvement of 1.3 percentage points.
	
\end{itemize}

Table \ref{table3} gives a detailed comparison of the proposed framework and comparative methods on Yago11k and Wikidata12k datasets. As can be seen from this table:
\begin{itemize}
	\item Different from ICEWS14, ICEWS05-15 and GDELT, the number of training samples is exponentially reduced in Yago11k and Wikidata12k. In this scenario, our proposed framework achieves consistent improvements. On both datasets, it refines the performance of all baseline TKG embedding models.
	
	\item Where the improved DE\_SimplE and TeLM methods outperform other models on ICEWS14 and ICEWS05-15, they also yield the best results on Yago11k and Wikidata12k while the BoxTE model does not obtain impressive performance on these small collections.
	
	\item On the Yago11k dataset, the proposed framework can improve the performance by an average of 5.1\% of MRR beyond the original methods, and gain an average improvement of 5.6\% of MRR on Wikidata12k.
	
	\item To summarize both tables, we observe that the method which achieves good results in MR does not perform well in the rest metrics. In other words, the model performance aspect reflected by MR is different from the other metrics. This is why  current studies rarely uses MR as a metric for measuring the performance of models.
\end{itemize}

\section{Discussion}
\label{section5}
In the following, we focus on analyzing and discussing the influence of some specific modules and parameters on model performance via ablation studies and demonstrate the effect of the proposed framework more graphically, include a visual representation of the embedding model and an illustration of some negative samples generated by random and generator sampling schemes.

\subsection{Model variants and ablation studies}
\begin{table}[htbp]
	\centering
	\caption{Evaluation results corresponding to different generator backbone methods on ICEWS14 dataset. The best result is formatted in boldface.}
	\renewcommand{\arraystretch}{1.4}
	\setlength\tabcolsep{2.5pt}
		\begin{tabular}{l|ccccc}
			\hline
			\multicolumn{1}{c|}{\multirow{2}{*}{Generator}} & MR              & MRR             & Hits@1\%        & Hits@3\%        & Hits@10\%       \\ \cline{2-6} 
			\multicolumn{1}{c|}{}                           & \multicolumn{5}{c}{F-TTransE}                                                           \\ \hline
			TTransE-based                                   & \textbf{167.64} & \textbf{0.2381} & 0.0481          & \textbf{0.3487} & \textbf{0.5966} \\
			TA\_DistMult-based                              & 183.47          & 0.2320          & 0.0454          & 0.3428          & 0.5817          \\
			DE\_SimplE-based                                & 191.29          & 0.2323          & \textbf{0.0645} & 0.3194          & 0.5619          \\ \hline
			& \multicolumn{5}{c}{F-TA\_DistMult}                                                      \\ \hline
			TTransE-based                                   & \textbf{173.20} & 0.3476          & \textbf{0.2161} & 0.3992          & 0.6259          \\
			TA\_DistMult-based                              & 185.19          & 0.3452          & 0.1920          & 0.3994          & \textbf{0.6380} \\
			DE\_SimplE-based                                & 187.58          & \textbf{0.3501} & 0.2112          & \textbf{0.4125} & 0.6352          \\ \hline
			& \multicolumn{5}{c}{F-DE\_SimplE}                                                        \\ \hline
			TTransE-based                                   & \textbf{296.79} & 0.4504          & 0.3237          & 0.4976          & 0.6753          \\
			TA\_DistMult-based                              & 312.94          & 0.4561          & 0.3378          & 0.5171          & \textbf{0.6929} \\
			DE\_SimplE-based                                & 304.42          & \textbf{0.4574} & \textbf{0.3395} & \textbf{0.5177} & 0.6904          \\ \hline
		\end{tabular}
\label{table4}
\end{table}

In the above, we employ a simple TTransE model as the generator, in this section, investigate the impact of generators with different TKGE methods on the overall performance of the framework. Table \ref{table4} illustrates the experimental results acquired via replacing the vanilla version of the generator with other TKGC models, such as TA\_DistMult and DE\_SimplE. We can see that when applying TTransE as the discriminator, a TTransE-based generator can most effectively improve the model performance, and a framework involving DE\_SimplE-based generators yields the best performance with DE\_SimplE discriminators. When the discriminator is TA\_DistMult or DE\_SimplE, we utilize the generator based on TA\_DistMult to obtain the best score on Hits@10\%. 

In general, it is contrary to our intuition that switching different TKGE generators would not significantly affect the performance of the proposed adversarial framework. A possible explanation for this observation might be that the goal of the generator is to construct plausible negative samples, requiring just enough representation capacity to calculate which entities are similar to the removed original entities in feature space to construct negative samples. However, an overpowered generator may cause the constructed samples to become too homogeneous, leading to issues such as mode collapse.

\begin{figure*}
	\subfigure[]{
		\begin{minipage}[htb]{0.23\linewidth}
			\centering
			\includegraphics[height=1.25in]{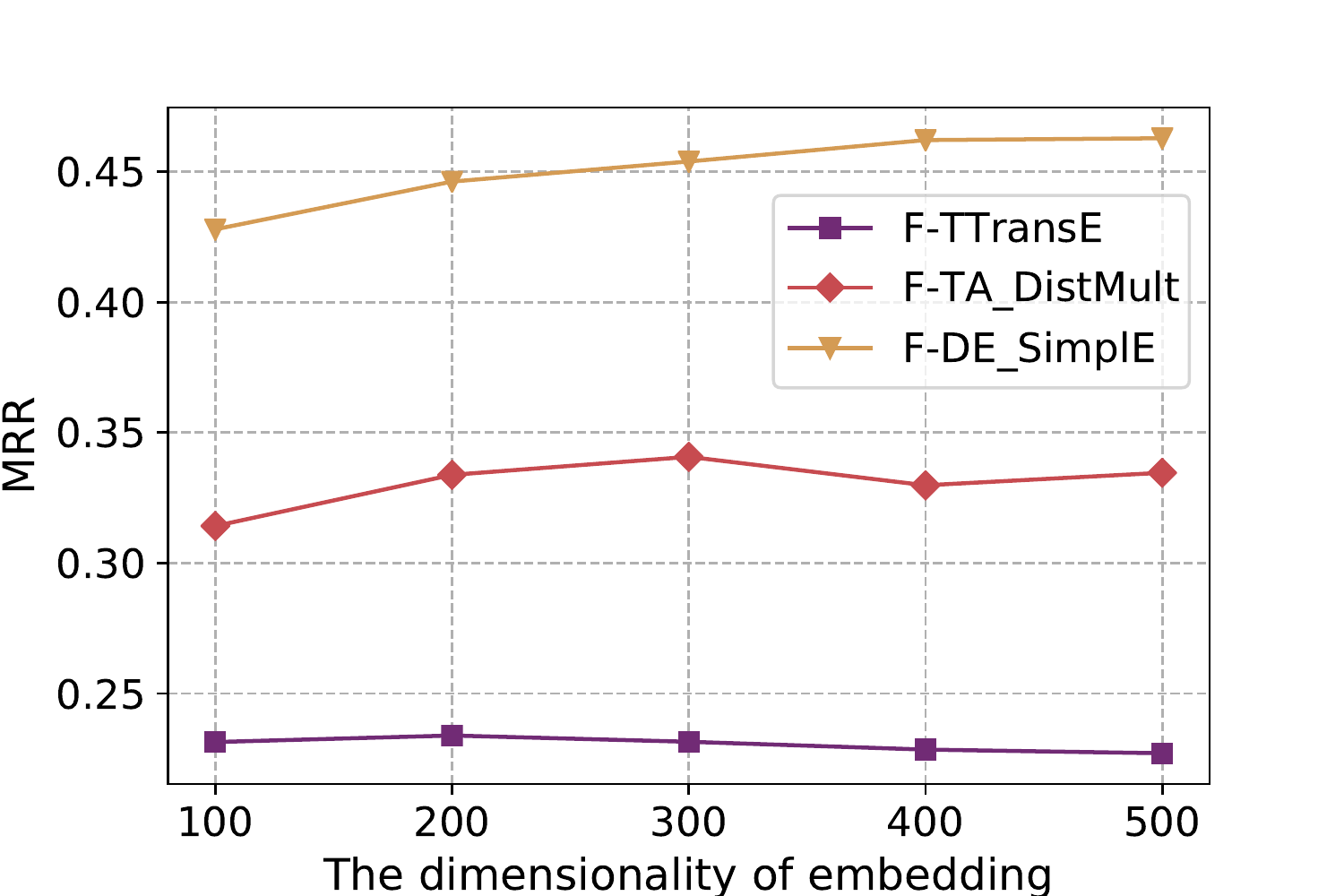}
		\end{minipage}
	}
	\subfigure[]{
		\begin{minipage}[htb]{0.23\linewidth}
			\centering
			\includegraphics[height=1.25in]{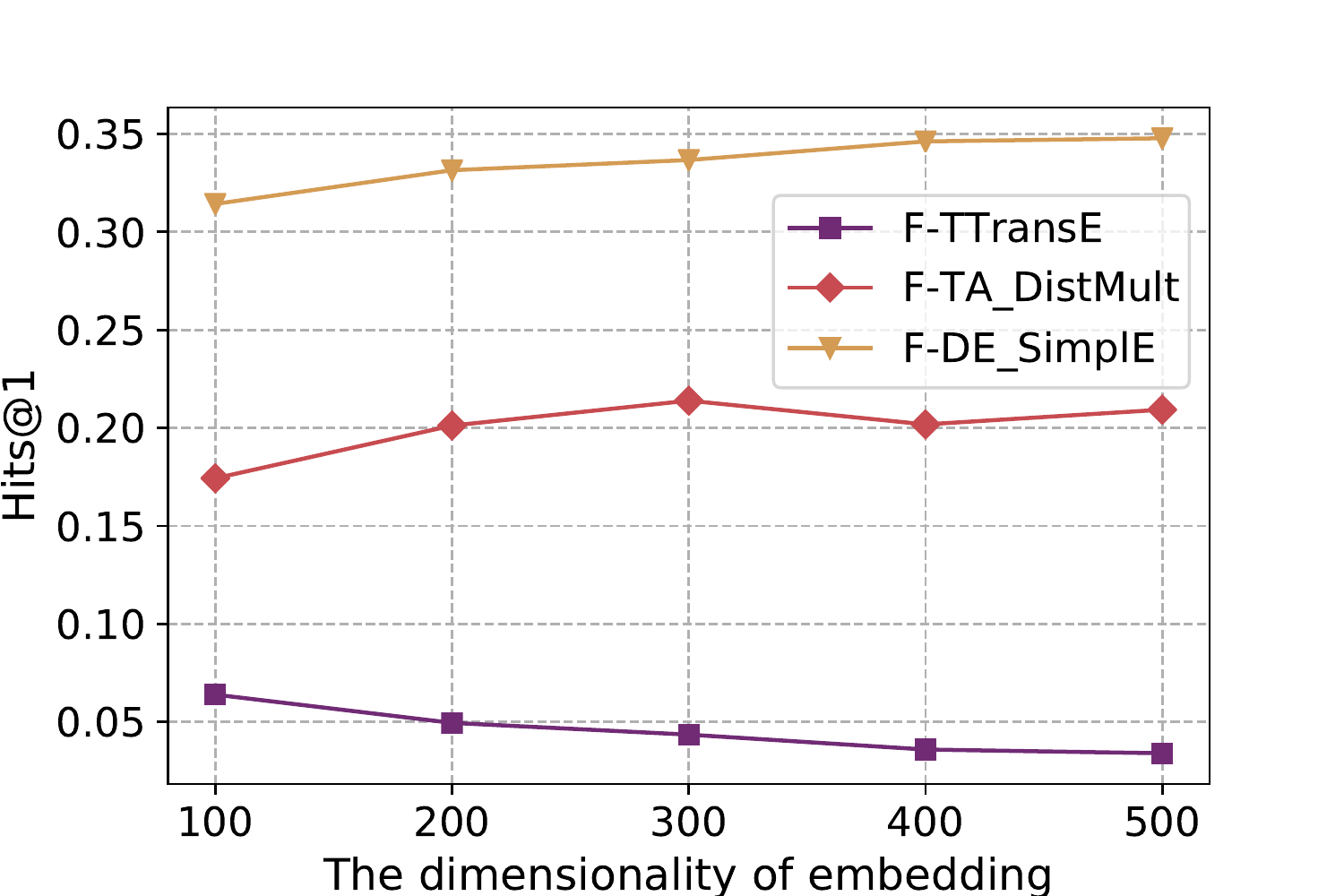}
		\end{minipage}
	}
	\subfigure[]{
		\begin{minipage}[htb]{0.23\linewidth}
			\centering
			\includegraphics[height=1.25in]{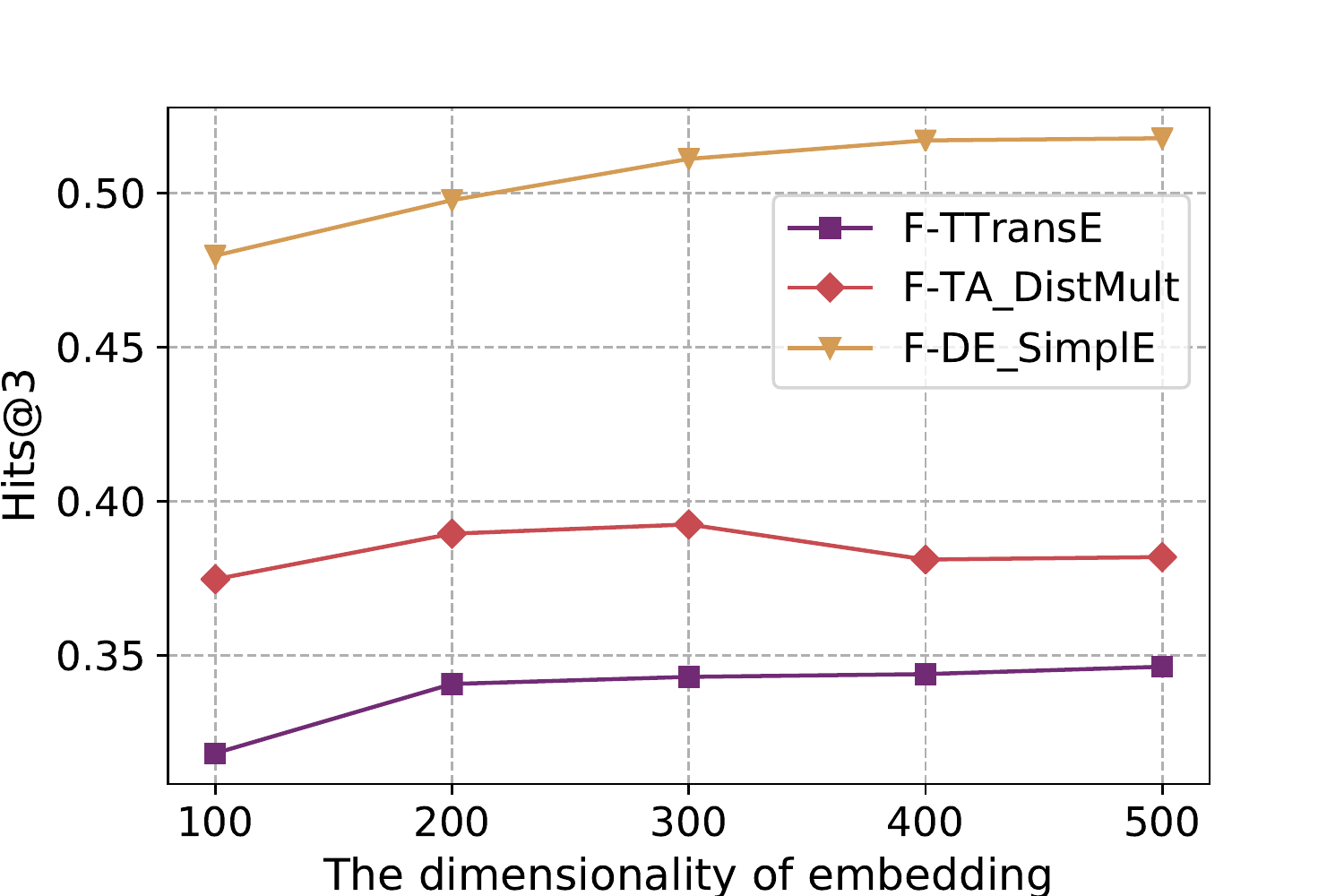}
		\end{minipage}
	}
	\subfigure[]{
		\begin{minipage}[htb]{0.23\linewidth}
			\centering
			\includegraphics[height=1.25in]{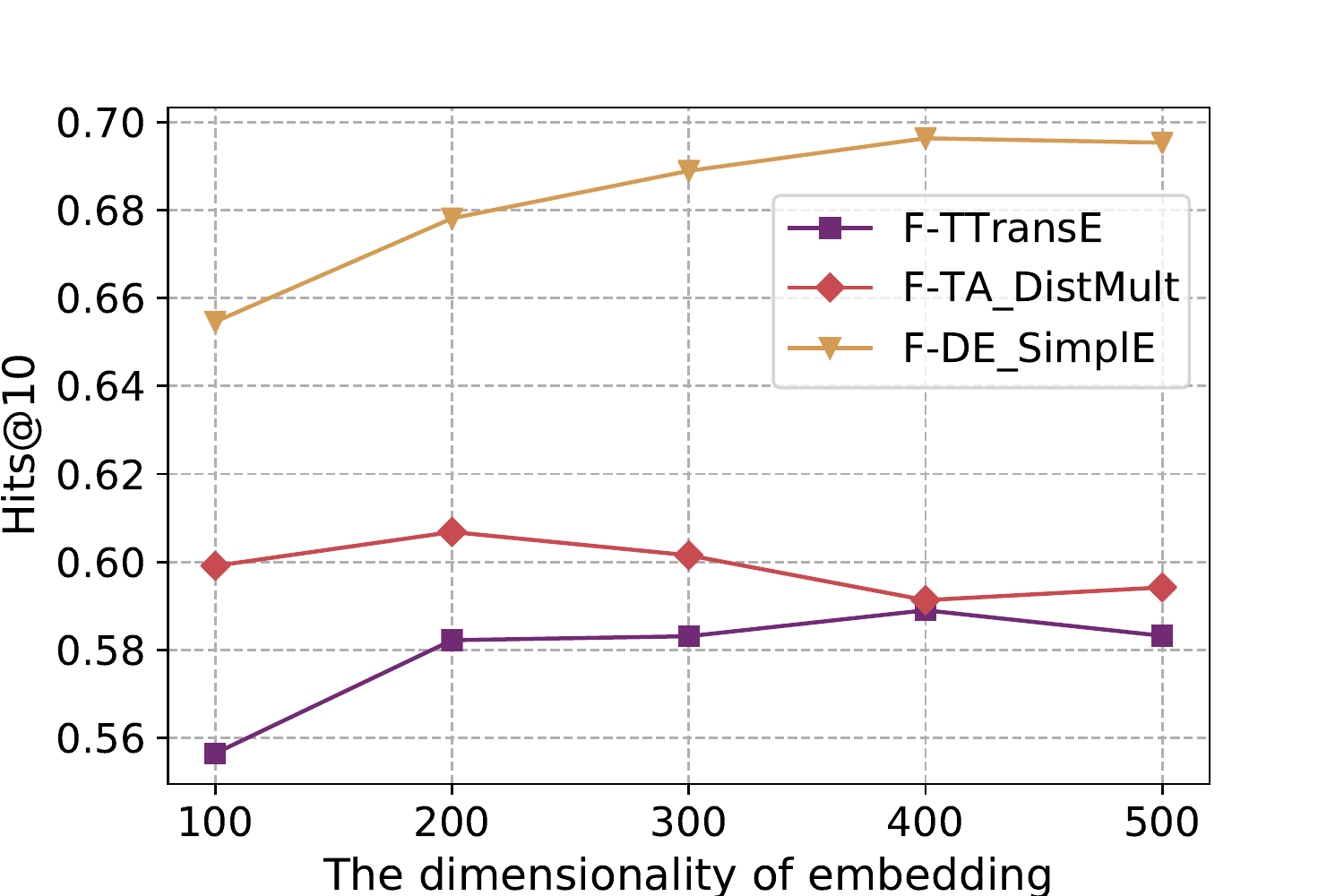}
		\end{minipage}
	}
	\caption{Results of F-TTransE, F-TA\_DistMlut and F-DE\_SimplE with different embedding dimensions on MRR, Hits@1, Hits@3 and Hits@10 metrics.} 
	\label{Figure4}
\end{figure*}

The embedding dimensionality is another significant hyper-parameter for each TKGE model. We conduct a contrast experiment where we train F-TTransE, F-TA\_DistMlut and F-DE\_SimplE on ICEWS14 for 100 epochs to compare the impact of different dimensions on the representation performance. Figure \ref{Figure4} plots the results of various TKGE model based frameworks with different embedding dimensions. As can be seen, although the performance of frameworks is indeed enhanced as the embedding dimension increases, the improvement is not significant, which is why we use 100 dimensions as the basis for our core experiments.

\subsection{Visualization}
In this section, we highlight the representation capabilities of the proposed framework in a qualitative manner via two visualization experiments: an illustration of negative samples constructed by random sampling versus the generator mode, and a diagram of TKG embedding vectors.

Traditional random negative samples are compared to generated ones in Table \ref{table5}. In this table, the quadruples in the first column are positive, the underlined \underline{entities} indicate that they would be replaced by other entities in the next two columns. The three replacement entities which intend to supersede the underlined one with random sampling are listed in the second column, and the entities generated by our framework are displayed in the third column. It is apparent from this table that the generator is capable of selecting more plausible entities as negative samples. For instance, given an authentic quadruple $(\underline{Mexico}, Make\ optimistic\ comment, Vietnam, 2014-12-12)$, the generator adopts three semantically relevant head entities to replace $Mexico$, i.e. $Iran, Japan$ and $Government-Italy$. All of these entities have similar attributes and can formally represent a country, making the constructed facts plausible and potentially deceptive.
\begin{table*}[]
	\caption{Some instances of negative samples constructed by random and generator sampling mode from the ICEWS14 dataset. All relations are signified by star ($\bigstar$).}
	\centering
	\renewcommand{\arraystretch}{1.3}
	\setlength\tabcolsep{12pt}
		\begin{tabular}{c|c|c}
			\hline
			\textbf{Positive quadruples}         & \textbf{Random sampling} & \textbf{Generator sampling}     \\ \hline
			\underline{Barack Obama}                         & Employee -- India        & Businessperson -- United States \\
			$\bigstar$ Make a visit (2014-04-22) & Middle East                       & John Kerry              \\
			Malaysia                  & Police -- Kenya                   & Yannis Stournaras       \\ \hline
			Presidential Candidate -- Argentina  & Ministry -- Nigeria      & Foreign Affairs -- South Korea  \\
			$\bigstar$ Consult (2014-10-09)      & France                            & Military -- Philippines \\
			\underline{Congress -- Argentina}     & Djibouti                          & China                   \\ \hline
			\underline{Mexico}                    & Member of the Judiciary -- Canada & Iran                    \\
			$\bigstar$ Make optimistic comment (2014-12-12) & Suleiman Abba            & Japan               \\
			Vietnam                   & Catherine Ashton                  & Government -- Italy     \\ \hline
	\end{tabular}
	\label{table5}
\end{table*}

\begin{figure*}
	\subfigure[]{
		\begin{minipage}[htb]{0.5\linewidth}
			\centering
			\includegraphics[height=2in]{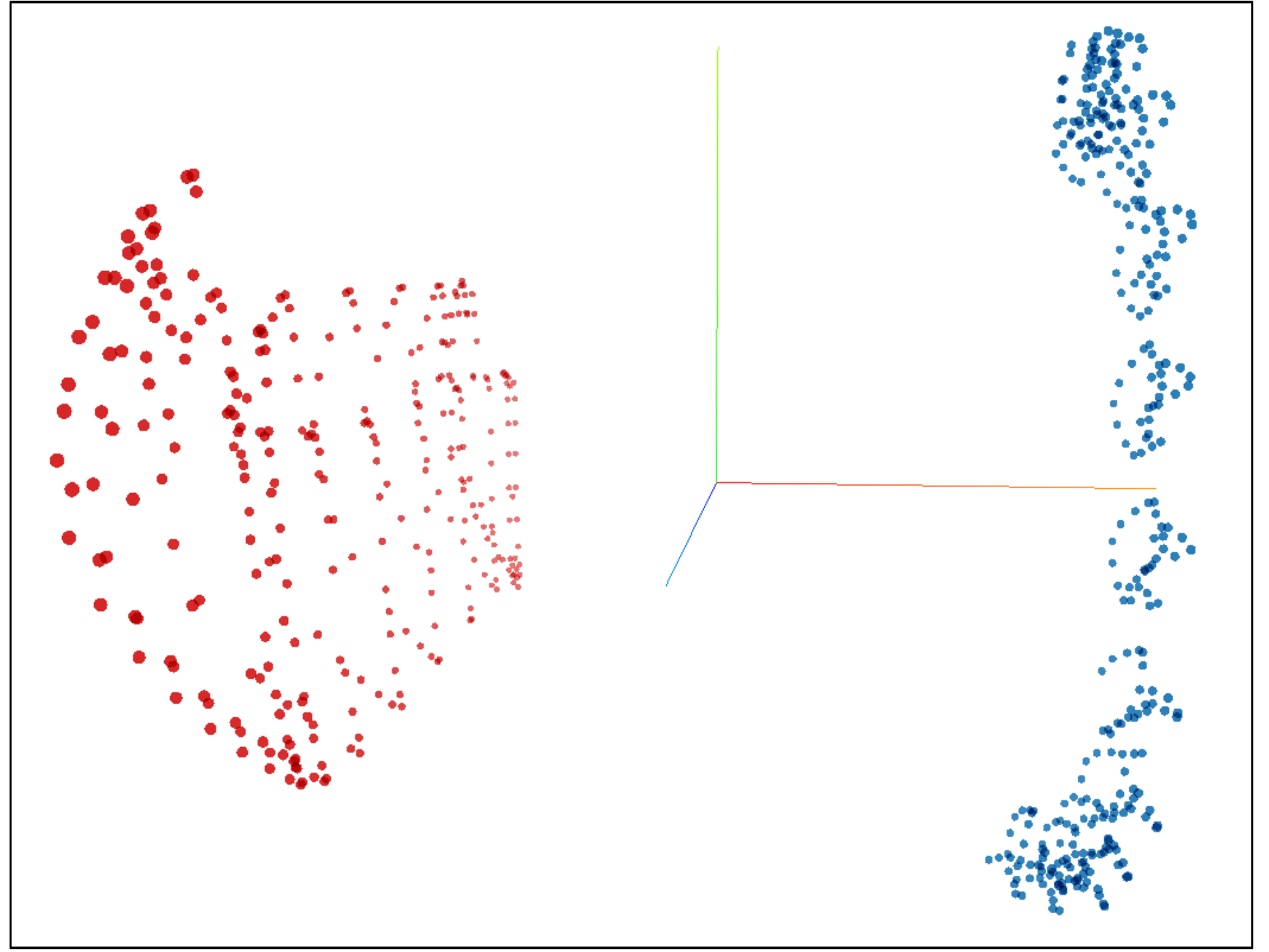}
		\end{minipage}
		\label{Figure5.1}
	}
	\hfill
	\subfigure[]{
		\begin{minipage}[htb]{0.5\linewidth}
			\centering
			\includegraphics[height=2in]{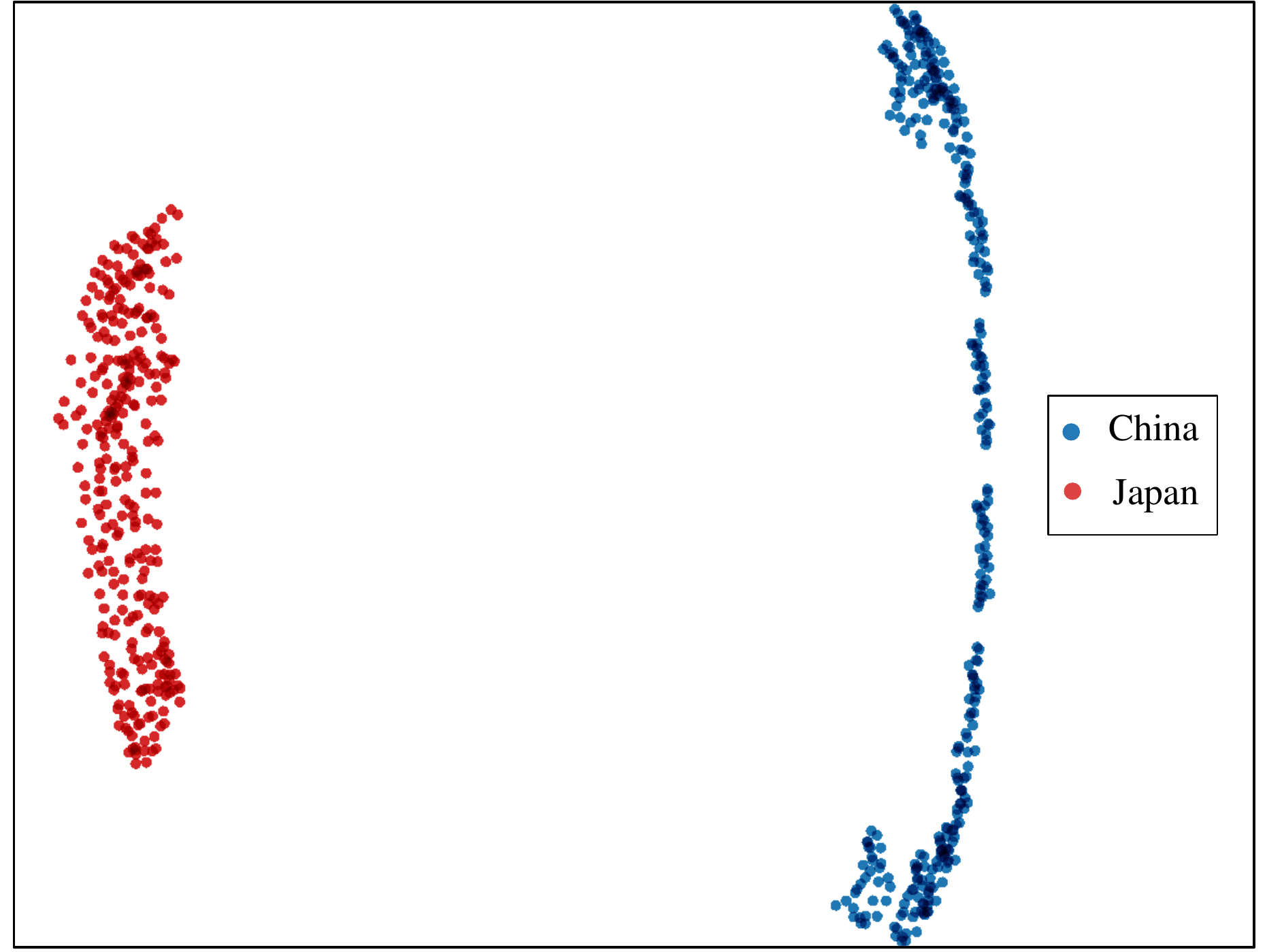}
		\end{minipage}
	\label{Figure5.2}
	}
	\vfill
	\subfigure[]{
		\begin{minipage}[htb]{0.5\linewidth}
			\centering
			\includegraphics[height=2in]{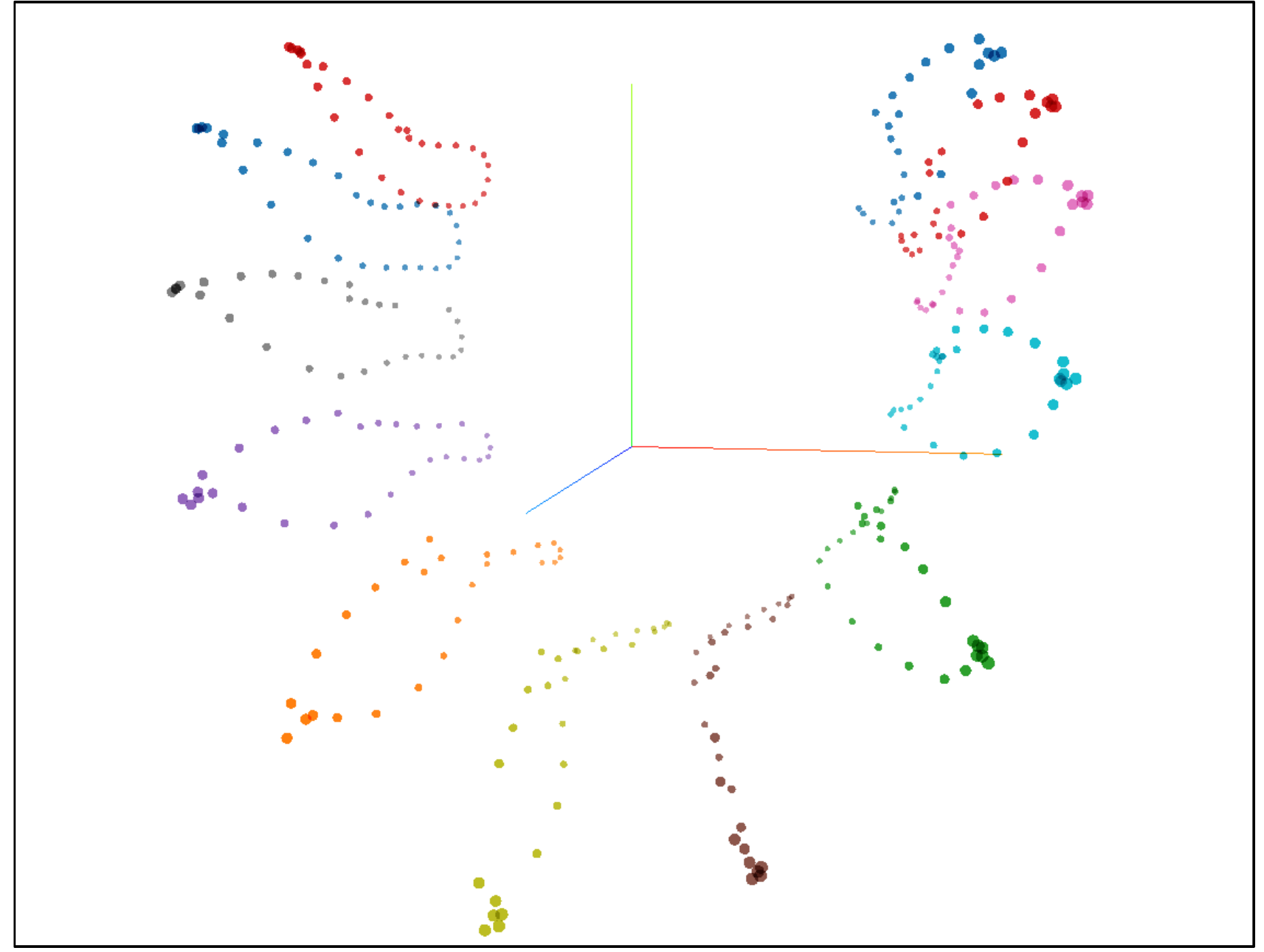}
		\end{minipage}
	\label{Figure5.3}
	}
	\hfill
	\subfigure[]{
		\begin{minipage}[htb]{0.5\linewidth}
			\centering
			\includegraphics[height=2in]{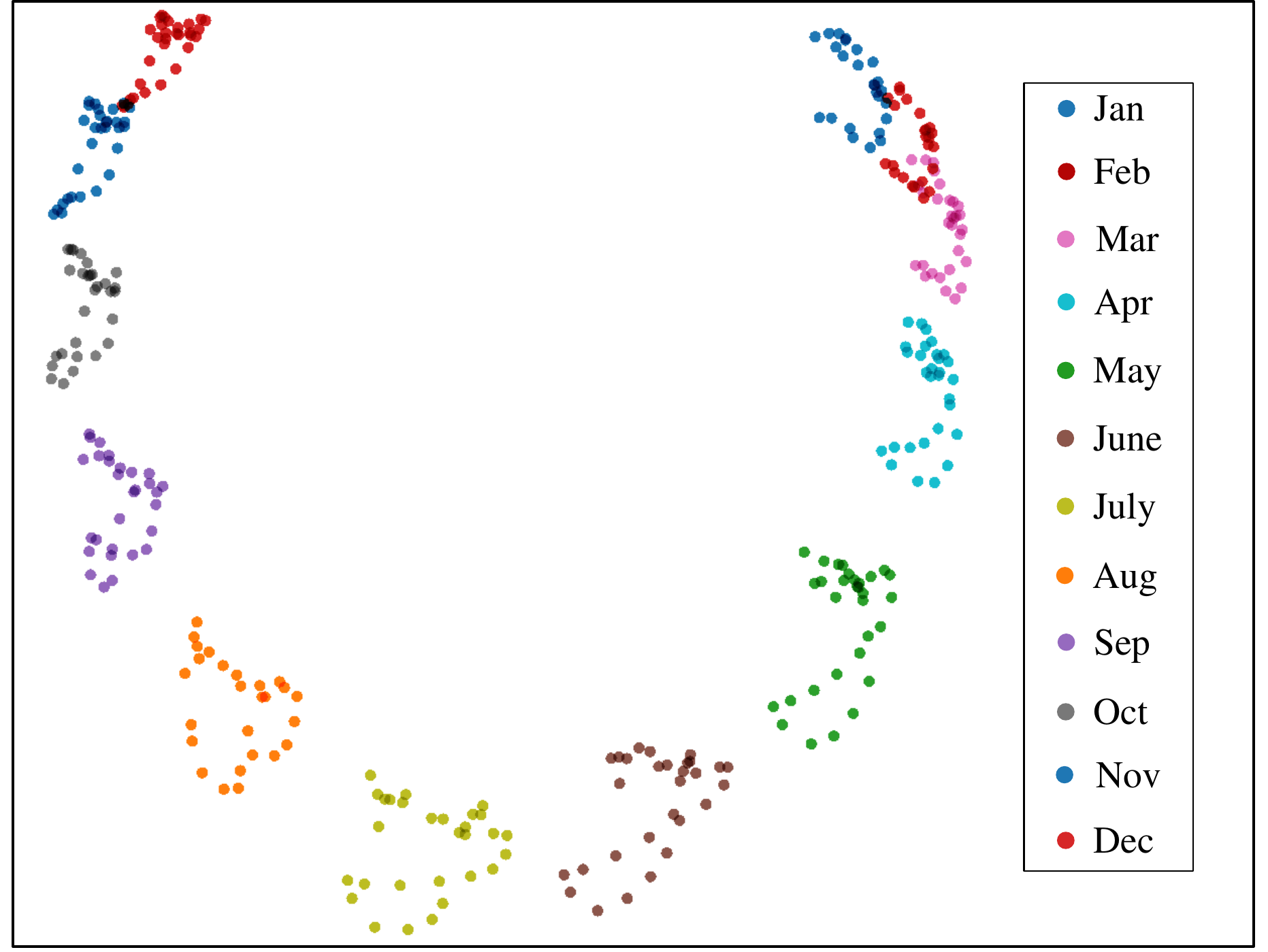}
		\end{minipage}
	\label{Figure5.4}
	}
	\vfill
	\caption{Diagram of TKG embedding vectors. (a) 3D overview of embedding vectors associated with China and Japan. (b) 2D overview of embedding vectors associated with China and Japan. (c) 3D overview of embedding vectors of China. (d) 2D overview of embedding vectors of China.}
	\label{Figure5}
\end{figure*}

When obtaining such high-quality negative quadruples by our proposed framework, we can train better TKGE models which have improved generalization and representation capabilities. As mentioned in Section \ref{section2}, TKGE approaches are similar to word embedding, following the basic translation invariance principle that entities with similar connotation should have similar representations. Dimensionality reduction via PCA is applied to project the trained entity vectors into a two and three-dimensional space to demonstrate whether they satisfy this principle.

Figure \ref{Figure5} shows a diagram of TKG embedding vectors after dimensionality reduction. We first acquire 343 and 313 temporal embedding vectors associated with entity $China$ and $Japan$ on the ICEWS14 dataset respectively, and then obtain a 3D and 2D overview of them via dimensionality reduction in Figure \ref{Figure5.1} and \ref{Figure5.2}. As these two figures show, the vectors of the two parts, $China$ and $Japan$, can be completely separated, while multiple time-series vectors belonging to the same entity can be tightly clustered together. This result illustrates that the model trained by our framework can make different semantic entities occupy different positions in the feature space, and also ensure that entities with the same semantic but different temporal information are close to and distinguished from each other, in accordance with the translational invariance principle. To further validate the representational effect of our framework, we perform a more fine-grained partitioning of the 343 time-series embedded entities about $China$, marking them with different colors by month, and use the same dimensionality reduction technique to yield their 3D and 2D overviews Figure \ref{Figure5.3} and \ref{Figure5.4}. What stands out in the figure is that these embedding vectors can be aggregated by month and are distinct from one another in feature space. This illustration also intuitively supports the effectiveness of the proposed framework from a qualitative perspective.

\section{Conclusions}
\label{section6}
This paper presents a new negative sampling strategy that improves the performance of arbitrary temporal knowledge graph embedding models. To this end, we present an adversarial learning approach framework based on Wasserstein distance. To evaluate the performance of our proposed framework, we construct detailed link prediction experiments. The results on five standard collections confirm that our adversarial learning approach can significantly improve the performance of all baseline TKGE models.

Compared to conventional TKG embedding models, this approach has several significant advantages. First, we introduce an adversarial learning framework to represent TKGs. The generator is utilized to produce more plausible entities as negative samples, and these negative quadruples are fed to the discriminator alongside authentic positive ones to improve temporal embedding model's performance. In order to solve the inherent issue of vanishing gradients on discrete data, we employ a Gumbel-Softmax relaxation and the Wasserstein distance for ensuring the entire closed-loop back propagation process of this framework. Most notably, the work presented here can be applied to refine the performance of most existing TKGE models without requiring major modifications.


\bibliographystyle{IEEEtran}

\end{document}